\documentclass[prb,aps,nobibnotes,showkeys,twocolumn]{revtex4}
\usepackage[dvips]{graphicx}
\usepackage{amssymb,amsfonts,amsmath}
\usepackage{bm}
\usepackage{dcolumn}
\usepackage[dvips]{color}
\begin{document}

\title{\centering\Large\bf Energetics and Kinetics of Primary Charge Separation in
  Bacterial Photosynthesis}
\author{David N.\ LeBard}
\author{Vitaliy Kapko}
\author{Dmitry V.\ Matyushov}
\affiliation{Center for Biological Physics, 
         Arizona State University, PO Box 871604, Tempe, AZ 85287-1604}

\date{\today}
\begin{abstract}
  We report the results of Molecular Dynamics (MD) simulations and
  formal modeling of the free energy surfaces and reaction rates of
  primary charge separation in the reaction center of
  \textit{Rhodobacter sphaeroides}. Two simulation protocols were used
  to produce MD trajectories. Standard force field potentials were
  employed in the first protocol. In the second protocol, the special
  pair was made polarizable to reproduce a high polarizability of its
  photoexcited state observed by Stark spectroscopy. The charge
  distribution between covalent and charge-transfer states of the
  special pair was dynamically adjusted during the simulation run. We
  found from both protocols that the breadth of electrostatic
  fluctuations of the protein/water environment far exceeds previous
  estimates resulting in about 1.6 eV reorganization energy of
  electron transfer in the first protocol and 2.5 eV in the second
  protocol. Most of these electrostatic fluctuations become
  dynamically frozen on the time-scale of primary charge separation
  resulting in much smaller solvation contributions to the activation
  barrier. While water dominates solvation thermodynamics on long
  observation times, protein emerges as the major thermal bath coupled
  to electron transfer on the picosecond time of the reaction. Marcus
  parabolas were obtained for the free energy surfaces of electron
  transfer by using the first protocol while a highly asymmetric
  surface was obtained in the second protocol. A non-ergodic
  formulation of the diffusion-reaction electron transfer kinetics has
  allowed us to reproduce the experimental results for both the
  temperature dependence of the rate and the non-exponential decay of
  the population of the photoexcited special pair.
\end{abstract}
\preprint{Submitted to \textit{J. Phys. Chem. B}}
\maketitle

\section{Introduction}
\label{sec:0}
The problem of bacterial photosynthesis has received enormous
attention from both experimental and theoretical
communities.\cite{Fleming:88,Hoff:97,Page:99,Warshel:01,Blankenship:02,Page:03}
Here, we consider only the first step in the sequence of electronic
transitions following the absorption of a visible photon by the
special pair of the reaction center, the primary charge separation.
The calculation of the rate of primary charge separation, which over
several decades of intense research has effectively become the
hydrogen molecule of bioenergetics, involves two components: the
electronic communication between the primary donor and acceptor
responsible for electron
tunneling\cite{Zhang:98,Ivashin:98,Nishioka:05} and the Franck-Condon
factor describing the probability of bringing the donor and acceptor
levels into resonance with each other.\cite{Bixon:95,BixonJortner:99}
Our paper is concerned with that latter part of the problem which we
dub as the energetics of primary charge
separation.\cite{Volk:98,Winter:03,Noy:06}

In addressing the issue of the energetics of charge separation, we
first want to dissect this complex problem into two, not necessarily
simpler, questions: (1) What is the importance of the structural
arrangement of the cofactors in the reaction center protein?  and (2)
What are the roles played by the protein and hydrating water in
activating electronic transitions? Each of these questions has
generated a significant amount of literature on its own, and we will
not be able to provide a comprehensive discussion of each topic,
focusing instead on our main goal, the factors affecting the free
energy of activation.

Since optical spectroscopy of bacteriochlorophyll cofactors can be
studied separately, the most intriguing question related to our
discussion is how the energetics of optical transitions and electron
transfer are affected when the cofactors are assembled within the
protein matrix. The notion often circulated in the
literature\cite{Blankenship:02} is that protein provides a
low-polarity environment lowering the free energy of solvation of
embedded cofactors compared to solvation in water. Even though this
statement is generally correct, we will show below that nuclear
solvation approaching the thermodynamic limit of infinite observation
(waiting) time is still quite significant for the electron transfer
dipole formed by difference occupation numbers (atomic charges) of the
electron in the donor and acceptor states. In particular, solvation of
the electron transfer dipole by water is not fully screened by the
protein and still makes about 1 eV. In addition, the protein matrix
cannot be really considered non-polar since there is a significant
contribution to the reorganization energy from the nuclear modes of the
protein. It turns out that the notion of weak nuclear solvation of
primary charge separation, required to explain the observed rates,
cannot fully rest on the thermodynamic arguments, and the dynamics of
the protein/water thermal bath need to be involved.

Solvation dynamics of optical chromophores in dense molecular solvents
have been actively studied in the past
decades.\cite{Rossky:94,Jimenez:94,Reynolds:96} The basic picture,
first discovered in numerical simulations\cite{Maroncelli:88} and
later confirmed by laboratory measurements,\cite{Jimenez:94} is that
the decay of the solvation correlation function (Stokes shift
correlation function, $S(t)$) involves two major components. The fast
Gaussian component is caused by ballistic motions of the solvent in
the first solvation shell of the solute (quasi-localized vibrations in
the case of a protein). The slow tail of $S(t)$ is related to
collective $\alpha$ relaxation mostly caused by relaxation of orientations
of molecular permanent dipoles (dielectric relaxation) and
quadrupoles.\cite{Reynolds:96} The notion of $\alpha$ relaxation, that is
the slowest relaxation on the microscopic scale, is not commonly
invoked in the discussion of high-temperature solvation dynamics of
small molecular dyes,\cite{FlemingWolynes:90} but becomes critical in
building a conceptual basis for understanding the solvation dynamics
of cofactors assembled within the hydrated protein.\cite{Fenimore:04}

\begin{figure}
  \centering
  \includegraphics*[width=7cm]{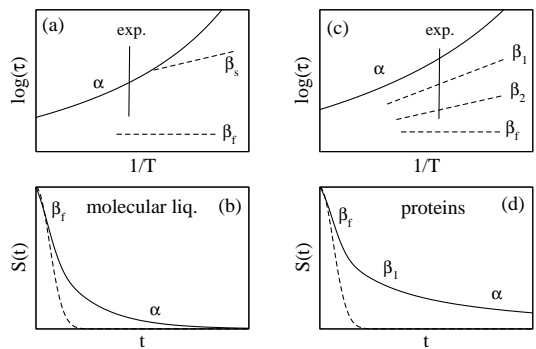}
  \caption{Relaxation times (a,c) and the Stokes shift dynamics (b,d) of
    the structural glass-formers (a,b) and proteins (c,d).  The
    vertical lines marked ``exp.'' denote the temperature at which the
    Stokes shift correlation function is recorded. The dashed lines in
    (b,d) show the fast Gaussian decay of $S(t)$.  
}
  \label{fig:1}
\end{figure}

Phenomenology developed for structural glass-formers\cite{Ediger:96}
helps to formulate the problem we are dealing with here. The typical
temperature dependence of the relaxation time of a polar molecular
liquid is shown in Figure \ref{fig:1}a. A high-temperature liquid has
two relaxation times: reorientations of molecular permanent dipoles
resulting in slow $\alpha$ relaxation and fast $\beta_f$ relaxation related to
collective anharmonic cage rattling. Correspondingly, the Stokes shift
correlation function has two components: fast Gaussian decay coupled
to $\beta_f$ molecular motions and a slow tail coupled to $\alpha$
motions. This latter component is often connected to dielectric
relaxation of the homogeneous solvent.\cite{Bagchi:91} When the liquid
is supercooled, the $\alpha$ component, which often becomes non-Arrhenius,
separates from the slow $\beta$ relaxation ($\beta_s$) characterized by the
Arrhenius temperature dependence\cite{Ediger:96} (Fig.\
\ref{fig:1}a). If all the components of the Stokes shift correlation
function could be resolved at that low temperature, three major parts,
corresponding to $\alpha$, $\beta_s$, and $\beta_f$ relaxation could have been
seen. It is this imaginary experiment, which is hard to realize in
molecular liquids,\cite{Ranko:00} that bears a close connection to
charge-transfer dynamics in proteins.

For proteins, as well as for most polymer glass-formers, $\alpha$ and $\beta$
relaxation are well separated in the temperature range of protein
stability.\cite{Ngai:00,Fenimore:04,Yang:07} In addition, the
secondary $\beta$ relaxation is typically split into several components
with increasingly faster dynamics accompanied by smaller activation
barriers (Figure \ref{fig:1}c). The rugged surface of the protein also
complicates the dynamics, and $\alpha$ relaxation is known to disappear from
the response of water in nano-confinement.\cite{Ngai:07} The coupling
of the transferred electron to different modes of the protein/water
solvent may vary, and it is \textit{a priori} not clear which mode
will dominate the solvation dynamics. However, one can clearly expect
Stokes shift dynamics to show at least three components including a
Gaussian decay due to $\beta_f$ modes, some subset of $\beta_s$ modes, and an
$\alpha$ relaxation (Figure \ref{fig:1}d).  The relative relaxation times
and weights of these modes in the overall Stokes shift correlation
function are critical for the energetics of charge transfer as we show
below.

\begin{figure}
  \centering
  \includegraphics*[width=7cm]{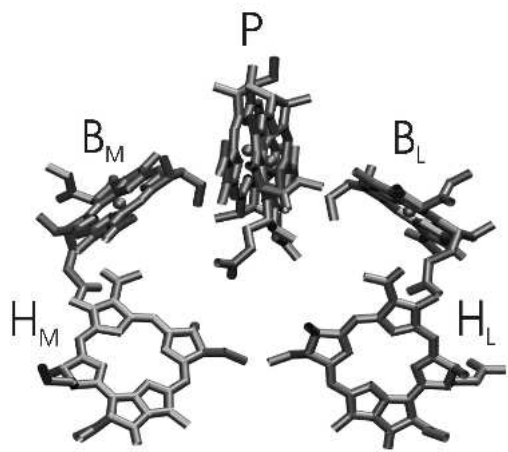}
  \caption{Schematic arrangement of cofactors in the bacterial
    reaction center. P is the special pair, B and H are monomeric
    bacteriochlorophylls and bacteriopheophytins, respectively. Electron
    transfer in wild-type reaction centers occurs almost exclusively
    along the L-branch of cofactors (subscript ``L''), while the
    M-branch (subscript ``M'') is mostly inactive. }
  \label{fig:2}
\end{figure}

The geometric arrangement of cofactors in the membrane protein of the
reaction center has been considered in the literature mostly from the
perspective of calculating the probability of electron tunneling
incorporated into the electron-transfer matrix element.\cite{Zhang:98}
Early studies considered the possibility of direct charge separation
from the special pair (P) to bacteriopheophytin (H$_L$) of the L
branch of monomeric chromophores via a super-exchange mechanism
involving nearby bacteriochlorophyll (B$_L$).\cite{Marchi:93,Bixon:95}
More recent
studies\cite{Schmidt:94,Ogrodnik:94,Holzwarth:96,Volk:98,Yakovlev:00}
have identified B$_L^-$ as an intermediate state in the sequence of
electron hops,\cite{Zinth:05} a slower process from P$^*$ to B$_L$
followed by a faster transition from B$_L$ to H$_L$. The energy level
of B$_L^-$ was placed between 331--450 cm$^{-1}$ (refs
\onlinecite{Schmidt:94,Holzwarth:96}) and 650--800 cm$^{-1}$ cm$^{-1}$
(ref \onlinecite{Roberts:01}) below the energy level of the excited
special pair P$^*$, favoring in both cases sequential over
superexchange transfer. In the present study, we will restrict our
attention to the first of two hops limiting our calculations to the
rate of transition from P$^*$ to B$_L$ (Figure \ref{fig:2}).

The role of the structural arrangement of the special pair in the
energetics of primary charge separation has attracted relatively
little attention (see, however, Warshel's work\cite{Warshel:80}). The
spectroscopy of the P$\to$P$^*$ excitation and of the primary pair
cation radical have been intensely
studied,\cite{Lathrop:94,Reimers:03} along with extensive modeling of
the energy transfer within the antenna complex and to the special
pair.\cite{Amerongen:00} The question we address here is somewhat
different. Given that the special pair has evolved within the reaction
center, we wonder if its particular structural arrangement makes any
significant impact on the activation barrier of primary charge
separation. Since the sandwich of two bacteriochlorophylls making P is
highly conserved in bacterial and plant
photosynthesis,\cite{Blankenship:02} it might have some other role in
the functionality of the reaction center aside from capturing the
excitation from the antenna complex. 

The motivation for posing this question is provided by Stark
experiments by Boxer and co-workers who showed a dramatic increase of
the polarizability of P upon photoexcitation.\cite{Middendorf:93} In
fact, the polarizability change of about 10$^3$ \AA$^3$ upon
photoexcitation\cite{Middendorf:93} places the special pair among the
most polarizable molecules known (carotenoids, also present in the
reaction center, make another group of champions). This remarkable
observation is combined here with our previous studies of electron
transfer in polarizable donor-acceptor
complexes,\cite{DMjpca:99,DMjacs:03,DMjpcb2:06} which showed that the
change in polarizability accompanying charge transfer results in
asymmetric, non-parabolic free energy surfaces for electron transfer.
Determining whether this polarization asymmetry can significantly
effect the activation barrier is one of the goals of this study.

In summary, by combining extensive Molecular Dynamics (MD) simulations
with formal modeling, we want to establish the basic ingredients
contributing to the activation barrier of primary charge separation.
The questions we address are the following: (1) What is the set of
primary nuclear modes (either protein or aqueous water/protein
interface) that promote transfer of an electron? (2) How to describe
the activation events happening on such a short reaction time? In
particular, we show that non-ergodic chemical kinetics is required in
this case to replace the standard Marcus picture based on equilibrium
distributions. (3) What is the effect of high polarizability of
the photoinduced special pair on the energetics of the transition?

\begin{figure}
  \centering
  \includegraphics*[width=7cm]{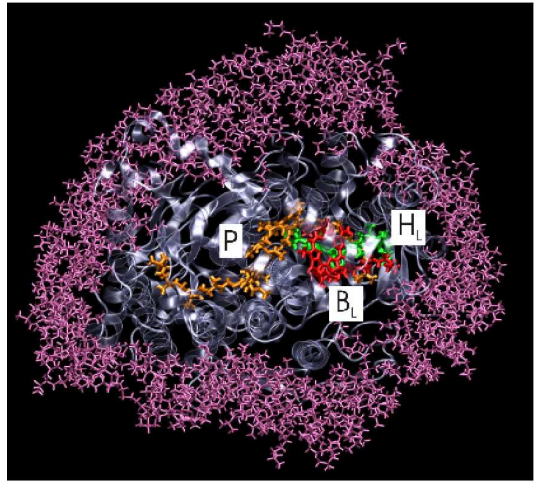}
  \caption{Reaction complex of \textit{Rhodobacter
      sphaeroides}.\cite{1pcr94} The protein (gray) is surrounded by
    the micelle of LDAO detergent molecules (purple). The electron is
    transferred in sequence from the photoexcited special pair (P,
    gold) to bacteriochlorophyll (B$_L$, red) followed by even faster
    transfer to bacteriopheophytin (H$_L$, green).  In MD simulations,
    the reaction complex is surrounded by 6 sodium ions, 30 NaCl
    pairs, and 10,506 water molecules which are not shown here. }
  \label{fig:3}
\end{figure}

We address the questions posed above by incorporating the Stokes shift
dynamics from MD simulations into a formal theory which we describe
first below. The results of the calculations presented next are tested
for consistency against experimental data. We use our simulation data
obtained at different temperatures to compare the calculated rates
with the results of Fleming \textit{et al}.\cite{Fleming:88} In
addition, the recently published data by Wang \textit{et
  al}.\cite{Wang:07} for the population decay of the photoexcited
state of the special pair in a number of mutants offer an opportunity
to use the dynamical electron transfer
models\cite{Sumi:86,Nadler:87,Zhu:91} to study the multiexponential
population decay. These experimental results are also analyzed here by
combining the Stokes shift dynamics from MD simulations with a formal
model of non-ergodic chemical kinetics. The picture that has emerged
from all this effort is summarized in the discussion section of this
paper.

\section{Basics of the Formalism}
\label{sec:2}
We approach the problem of calculating the rates of charge separation
by combining the input from MD simulations with analytical formalism.
Simulations of the reaction center of \textit{Rhodobacter
  sphaeroides}\cite{1pcr94} were carried out using Amber
8.0.\cite{amber8} We have followed the procedure first suggested by
Ceccarelli and Marchi\cite{Ceccarelli1:03} in which the reaction
center is surrounded by the micelle of detergent (lauryl dimethyl
amino oxide, LDAO) molecules mimicking the hydrophobic membrane, and
also closely matching the experimental setup\cite{Kirmaier:03} for
photochemical studies of bacterial photosynthesis. The structure of
the reaction center surrounded by LDAO molecules is shown in Figure
\ref{fig:3}. The details of the simulation protocol are provided in
Appendix \ref{secA} and the charging scheme of the cofactors and
the protein/water solvent is outlined in Appendix \ref{secB}.

\subsection{Energetics of Primary Charge Separation}
\label{sec:2-1}
Electron transfer is a tunneling event realized, in the
Born-Oppenheimer approximation, at the resonance of the electronic
donor and acceptor energies. The gap between the electronic energies
of the donor and acceptor states, $\Delta E$, makes the one-dimensional
reaction coordinate $X=\Delta E$ that incorporates the whole manifold of
possible nuclear modes affecting the electronic
transition.\cite{DMacc:07} Because many nuclear motions affect the
donor and acceptor states in dense condensed media, the fluctuations
of the stochastic variable $\Delta E$ are often well represented by
the Gaussian statistics.\cite{Kuharski:88} Therefore, the probability of
reaching zero energy gap $\Delta E=0$, when electron tunneling takes place,
is given by the Gaussian probability
\begin{equation}
  \label{eq:1}
  P(\Delta E=0) \propto e^{-\langle\Delta E\rangle^2/2C_X(0)}
\end{equation}
Here, the variance $C_X(0)=\langle (\delta X)^2\rangle$, $\delta X = \Delta E - \langle\Delta E\rangle$ is equal
to the $t=0$ value of the time self-correlation function of the energy
gap $X(t)=\Delta E(t)$
\begin{equation}
  \label{eq:13}
  C_X(t)=\langle\delta X(t) \delta X(0)\rangle 
\end{equation}
and the brackets stand for an ensemble average.

If one multiplies the probability of reaching the resonance with the
frequency $\omega_e$ of electron tunneling between the donor and acceptor
electronic levels, one arrives at the Marcus-Levich equation for the
electron transfer rate\cite{MarcusSutin}
\begin{equation}
  \label{eq:2}
  k_{\text{ET}} = \omega_e  e^{-\langle\Delta E\rangle^2/2C_X(0)} 
\end{equation}
In the case of non-adiabatic electron transfer considered here (weak
electronic coupling between the donor and acceptor), the electronic
tunneling frequency is given in terms of the electron transfer matrix
element $V$ by the following equation:
\begin{equation}
  \label{eq:3}
  \omega_e = \sqrt{2\pi/C_X(0)} (V^2/\hbar) 
\end{equation}

Equation \ref{eq:2} is quite general and is limited only by the assumption
of Gaussian fluctuations of the energy gap. In order to make it
practical, one needs to connect the average energy gap $\langle\Delta E\rangle $ and
the variance $C_X(0)$ to physical interactions present in the system made
of the donor-acceptor complex and a thermal bath of nuclear degrees
of freedom coupled to the transferred electron. Despite the obvious
complexity of the system, a generally applicable separation of the average
energy gap into three components is possible: the gas-phase gap $\Delta
E^{\text{gas}}$, the shift by non-polar interaction potentials $\Delta
E^{\text{ind}}$, and the shift by Coulomb interactions between the permanent
partial charges of the solute and the solvent, $\Delta E^{\text{C}}$
\begin{equation}
  \label{eq:4}
  \langle \Delta E\rangle = \Delta E^{\text{gas}} + \Delta E^{\text{ind}} + \Delta E^{\text{C}} 
\end{equation} 
The gas-phase energy gap $\Delta E^{\text{gas}}$ is the difference between
the ionization potential of the donor and the electron affinity of the
acceptor in the gas phase. The two other components represent the
interaction with the protein/water solvent and can thus be combined
into a solvent-induced (subscript ``s'') shift
\begin{equation}
  \label{eq:25}
  \Delta E_s =  \Delta E^{\text{ind}} + \Delta E^{\text{C}} 
\end{equation}

The separation of the average energy gap into a non-polar and Coulomb
part is in fact related to the separation of time-scales first
discussed in early work on polarons in solids by
Pekar,\cite{Pekar:46,Pekar:63} Fr{\"o}hlich,\cite{Frohlich:54} and
Feynman.\cite{Feynman:55} The fast electronic degrees of the solvent
(which is composed of protein, detergent, and water in our problem)
result in instantaneous equilibration of the transferred electron by
induction and dispersion (London) forces. For our present application,
the former is more significant (superscript ``ind'' in eq \ref{eq:4}
and throughout below) and we therefore explicitly consider this
component. The last term in eq \ref{eq:4}, related to Coulomb
interactions, fluctuates due to slow molecular motions of molecular
rotations and translations. This term is often described in the
electron transfer literature by the coupling of the electric field of
the donor-acceptor complex to the inertial dipolar
polarization,\cite{Pekar:63} which we consider after the induction
component.

The induction forces are produced by polarizing the medium by the
electric field of the donor-acceptor complex. If atoms and/or
molecules of the medium carry polarizabilities $\alpha_j$, the induction
energy is the sum of polarization free energies of all such
polarizable groups located at positions $\mathbf{r}_j$. The induction
shift of the average energy gap is then given by the change in the
polarization free energy caused by changing the electric field of the
donor-acceptor complex
\begin{equation}
  \label{eq:12}
   \Delta E^{\text{ind}} = -\left\langle \sum_j (\alpha_j/2)
     \left[E_{02}^2(\mathbf{r}_j) - E_{01}^2(\mathbf{r}_j)\right]\right\rangle 
\end{equation}
This component of the energy gap is often not given adequate attention
in the electron transfer literature, even though it can be quite
significant\cite{DMjcp:95,DMjcp2:05} as we show below. The induction
shift also depends on temperature for constant-pressure experiments
because of thermal expansion, and this fact needs to be included in
the modeling of temperature-dependent reaction rates. Even though the
induction potential is established instantaneously by induced
electronic dipoles, the interaction energy is modulated by nuclear
motions of the solvent (water and protein) producing a non-zero
component in the Gaussian distribution width (see below).

The Coulomb part of the donor-acceptor energy gap has received the
most attention over several decades of the theory development, and we
will briefly set up the stage for our treatment of this part of the
problem here. The linear response approximation, either in terms of
the electrostatic interaction with the medium dipolar
polarization\cite{Marcus:56} or in terms of partial atomic
charges,\cite{Raineri:99} has mostly been used as the basis for theory
development. In the former case, one considers the polarization of the
solvent by the electric field of the solute $\mathbf{E}_{01}$ in the
initial electron transfer state. This equilibrium polarization
$\mathbf{P}_{\text{eq}}(\mathbf{r})$ at point $\mathbf{r}$ within the
solvent is connected to the field $\mathbf{E}_{01}(\mathbf{r}')$ at
point $\mathbf{r}'$ by generally a nonlocal response function
$\bm{\chi}(\mathbf{r},\mathbf{r}')$\cite{Raineri:99,Kornyshev:96,DMcp:06}
\begin{equation}
  \label{eq:5}
  \mathbf{P}_{\text{eq}}(\mathbf{r}) = 
     \bm{\chi}(\mathbf{r},\mathbf{r}')*\mathbf{E}_{01}(\mathbf{r}')
\end{equation}
where the asterisk denotes volume integration over the variable
$\mathbf{r}'$ and tensor contraction over the Cartesian components of
the field with the corresponding components of the 2-rank tensor
$\bm{\chi}$.

Equation \ref{eq:5} for the solvent polarization induced by the solute
typically appears in theories of linear solvation in homogeneous
liquids. In contrast, the protein matrix itself and the protein/water
interface are inhomogeneous with a possibility of generating a
polarization field $\mathbf{P}'_{\text{eq}}$ unrelated to the electric
field of the cofactors. This polarization will create an additional
inhomogeneous component of the vertical shift $\Delta E_{\text{inh}}^C$
that cannot be calculated from the linear response approximation.
The inertial (nuclear) polarization field
$\mathbf{P}'_{\text{eq}}+\mathbf{P}_{\text{eq}}$ does not change on
the time-scale of electronic transition and creates a shift of the
donor-acceptor energy gap by the amount determined by the change in
electron's electric field $\Delta \mathbf{E}_0$ upon the transition:
\begin{equation}
  \label{eq:6}
  \Delta E^{\text{C}} = -\left(\mathbf{P}_{\text{eq}}'+\mathbf{P}_{\text{eq}}\right)*\Delta\mathbf{E}_0
                 = \Delta E^{\text{C}}_{\text{inh}} - \mathbf{E}_{01}*\bm{\chi}*\Delta\mathbf{E}_0
\end{equation}

In the original Marcus formulation,\cite{MarcusSutin} the average
vertical energy gap was separated into the Coulomb reorganization
energy $\lambda^{\text{C}}$ and the Coulomb part of the Gibbs energy of the reaction,
$\Delta G^{\text{C}}$. By using the identity $\mathbf{E}_{01}=\mathbf{\bar E}_0 - \Delta
\mathbf{E}_0/2$, $\mathbf{\bar E}_0=
(\mathbf{E}_{01}+\mathbf{E}_{02})/2$ in eq \ref{eq:6} one gets
\begin{equation}
  \label{eq:7}
   \Delta E^{\text{C}} = \lambda^{\text{C}} + \Delta G^{\text{C}} 
\end{equation}
where
\begin{equation}
  \label{eq:8}
   \lambda^{\text{C}} = (1/2) \Delta\mathbf{E}_0*\bm{\chi}*\Delta\mathbf{E}_0
\end{equation}
and
\begin{equation}
  \label{eq:20}
     \Delta G^{\text{C}} = \Delta E^{\text{C}}_{\text{inh}} - \mathbf{\bar E}_0*\bm{\chi}*\Delta\mathbf{E}_0
\end{equation}
The Coulomb part of the Gibbs energy then combines with the gas-phase
gap and the induction shift to make the overall reaction Gibbs energy
\begin{equation}
  \label{eq:9}
  \Delta G = \Delta E^{\text{gas}} + \Delta E^{\text{ind}} + \Delta G^{\text{C}} 
\end{equation}

Combined together, eqs \ref{eq:4}, \ref{eq:7}, and \ref{eq:9} lead to
the standard Marcus relation for the vertical average energy gap 
\begin{equation}
  \label{eq:10}
  \langle \Delta E \rangle = \Delta G + \lambda^{\text{C}}
\end{equation}
The separation of the average energy gap into the equilibrium Gibbs
energy and reorganization energy components makes sense when the
former can be measured separately. In spectroscopy, the average gap is
given by the maximum of the corresponding spectroscopic band (or, more
precisely, by the first spectral moment) and this separation is not
necessary. Likewise, the average energy gap is directly accessible
from MD simulations, so the formulation in terms of the average energy
gap is also more convenient from the simulation perspective.  Even
more importantly, the Gibbs energy of the reaction loses its direct
connection to equilibrium thermodynamics in non-ergodic reaction
kinetics, which we formulate and apply to the calculation of the rates
below. In this framework, the formulation of electron transfer
thermodynamics in terms of the first and second cumulants of the
donor-acceptor energy gap is the only formal approach to the problem
available at the moment.

The Gaussian width, $C_X(0)$ in eq \ref{eq:2}, generally needs a
separate determination. It is calculated as the variance of the sum of
all solute-solvent interaction potentials affecting the energy of the
transferred electron. The problem is simplified for the Coulomb
interactions. These are long-ranged and are typically well described
by the linear response approximation. Therefore, the high-temperature
limit of the fluctuation dissipation theorem\cite{Landau5} applies to
the Coulomb part on nuclear fluctuations with the resulting
factorization of $C_X(0)$ into temperature and reorganization
energy\cite{MarcusSutin,Ovchinnikov:69}
\begin{equation}
  \label{eq:46}
  C_X(0) = 2k_{\text{B}}T \lambda_s
\end{equation}
A significant simplification of this route is achieved through the
fact that the variance is determined in terms of the same response
function as the one used for the Coulomb part of the average energy
gap (eq \ref{eq:6}), thus reducing the number of independent response
functions required by the theory.

This procedure does not apply to short-range induction forces which do
not follow the macroscopic fluctuation-dissipation theorem; the
calculation of their first and second cumulants requires microscopic
response functions.\cite{DMjcp:95} The main consequence is that the
induction component does not factorize into temperature and a weakly
temperature-dependent energy parameter. The result is a generally
non-Arrhenius form of the rate constant\cite{DMacc:07} in eq
\ref{eq:2} in which the variance can be written as
\begin{equation}
  \label{eq:19}
  C_X(0) \simeq  2k_{\text{B}}T \lambda^{\text{C}} + C^{\text{ind}}(0)
\end{equation}
Since the Coulomb and induction interaction sum up in the energy gap,
a cross term needs to be taken into account, and we have included it
into $C^{\text{ind}}(0)$ as follows
\begin{equation}
  \label{eq:22}
  C^{\text{ind}}(0) = \langle (\delta E^{\text{ind}})^2 \rangle + 2 \langle\delta E^{\text{C}} \delta E^{\text{ind}} \rangle 
\end{equation}

Despite these complications which take away the solid foundation
behind factoring the variance into the temperature and energy
components,\cite{Landau5,Ovchinnikov:69} we will follow the
established tradition and define the solvent reorganization energy as
as the sum of induction and Coulomb terms (\textit{cf}.\ to eq \ref{eq:25})
\begin{equation}
  \label{eq:48}
    \lambda_s = \lambda^{\text{ind}} +  \lambda^{\text{C}} 
\end{equation}
where 
\begin{equation}
  \label{eq:49}
          \lambda^{\text{ind}}= C^{\text{ind}}(0)/(2k_{\text{B}}T)  
\end{equation}

The probability of electron transfer can be affected by intramolecular
vibrations of the solute.\cite{BixonJortner:99} These can be added to
the formalism outlined here by summing up probabilities of transitions
between separate vibronic channels. These transitions are known to
significantly affect the transition probability in the inverted
region, $\langle \Delta E \rangle <0$, but can be neglected for transitions in the
normal region, $\langle \Delta E \rangle > 0$, considered here. An extension to the
former case is well developed in the literature\cite{BixonJortner:99}
and does not pose fundamental difficulties.

\subsection{Stokes shift dynamics }
\label{sec:2-2}
The characteristic timescales of nuclear fluctuations affecting charge
transfer can be extracted from the time correlation function in eq
\ref{eq:13}, or from its normalized value known as the Stokes shift
correlation function
\begin{equation}
  \label{eq:14}
  S(t) = C_X(t)/C_X(0) 
\end{equation}
As mentioned above, the typical shape of $S(t)$ in complex condensed
media includes a fast Gaussian component and a multi-exponential (or
stretched-exponential) tail. A two-exponential tail is used to fit our
simulation results with $C_X(t)$ in the form
\begin{equation}
  \label{eq:15}
  C_X(t) = C^{\text{ind}}(t)+ C^{\text{C}}(t)
\end{equation}
where
\begin{equation}
  \label{eq:50}
C^{\text{C}} (t) =   2k_{\text{B}}T\left[\lambda_G^{\text{C}}  e^{-(t/ \tau_G)^2} + 
           \lambda_1^{\text{C}} e^{-t/ \tau_1} + \lambda_2^{\text{C}} e^{-t/ \tau_2}\right]
\end{equation}
Here, $\tau_G$ is the relaxation time of the Gaussian decay and $\tau_1$ and
$\tau_2$ are two exponential relaxation times. In addition, $\lambda_G$ and
$\lambda_i$ are the corresponding reorganization energy components such that
$\lambda^{\text{C}} =\lambda_G^{\text{C}} + \lambda_1^{\text{C}} +\lambda_2^{\text{C}}$.

\subsection{Non-ergodic activation kinetics}
\label{sec:2-3}
The arguments presented in sec \ref{sec:2-1} are based on equilibrium
statistical mechanics representing the components of the activation
barrier as equilibrium (free) energies. This formulation in fact
assumes a certain separation of time-scales, that is the time of the
reaction $\tau_{\text{ET}}= k_{\text{ET}}^{-1}$ must be much longer than
all relaxation times ($\tau_G$, $\tau_i$, etc.) of the nuclear modes coupled
to the transferred electron. This assumption certainly breaks down for
our problem combining the extremely short time of natural primary
charge separation (ca.\ 3 ps) with the disperse relaxation spectrum of
the protein/water solvent.\cite{Holzwarth:96,Parson:98,Winter:03} What
we face here is the obvious case of ergodicity
breaking\cite{Palmer:82} of the nuclear fluctuations involved in the
reaction activation, which raises the question of how to approach the
calculation of the reaction rates.

The Stokes shift correlation function provides a consistent approach
to formulate the kinetics of non-ergodic electron transfer. We first
note that the equilibrium linear response function $\bm{\chi}$,
introduced in Sec.\ \ref{sec:2-1} in the direct space domain, can be
extended to the time domain to cover the time correlation functions of
the energy gap fluctuations. The equilibrium ensemble average
producing the solvent response component of the average energy gap
(\textit{cf}.\ to eq \ref{eq:6}) can then be given as a frequency integral of the
Fourier transform $\bm{\chi}(\omega)$
\begin{equation}
  \label{eq:16}
  \Delta E^{\text{C}}_r = -2\int_0^{\infty} d\omega \mathbf{E}_{01}*\bm{\chi}(\omega)*\Delta\mathbf{E}  
\end{equation}

This representation offers a systematic approach to calculating the
non-ergodic solvent response. The integral in eq \ref{eq:16} is over
all possible frequencies of nuclear motions, implying that all of them
contribute to the average. In fact, the time-scale of the reaction
$\tau_{\text{ET}}$ limits the frequency spectrum only by those
frequencies that are higher than the rate of the reaction
$k_{\text{ET}}$. The non-ergodic energy gap thus becomes
\begin{equation}
  \label{eq:17}
    \Delta E^{\text{C}}_r(k_{\text{ET}}) = 
 -2\int_{k_{\text{ET}}}^{\infty} d\omega \mathbf{E}_{01}*\bm{\chi}(\omega)*\Delta\mathbf{E}_0
\end{equation}
Along the same lines, the non-ergodic reorganization energy can be
defined by using the same step-wise frequency filter:
\begin{equation}
  \label{eq:33}
  \lambda^{\text{C}}(k_{\text{ET}}) = \int_{k_{\text{ET}}}^{\infty} d\omega \Delta\mathbf{E}_0 *
                         \bm{\chi}(\omega)*\Delta\mathbf{E}_0
\end{equation}
An alternative representation is through the Fourier transform of the
Stokes shift correlation function
\begin{equation}
  \label{eq:23}
  C_X^{\text{C}}(\omega) = \int_{-\infty}^{\infty}e^{i\omega t} C_X^{\text{C}}(t) dt/(2\pi)
\end{equation}
as follows\cite{DMjcp2:06}
\begin{equation}
  \label{eq:18}
  \lambda^{\text{C}}(k_{\text{ET}})=   \beta \int_{k_{\text{ET}}}^{\infty} C_X^{\text{C}}(\omega) d \omega
\end{equation}
where $\beta = 1/(k_{\text{B}}T)$.  Equations \ref{eq:17}--\ref{eq:18}
suggest that $\lambda^{\text{C}}(k_{\text{ET}})$ can be obtained from the
Stokes shift correlation function calculated from MD trajectories
while a formal theory is required for $\bm{\chi}(\omega)$ to determine $\Delta
E_r^{\text{C}}(k_{\text{ET}})$.

The notion that the parameters entering the activation barrier become
functions of the electron transfer rate creates the necessity to
consider the calculation of the rate constant as a self-consistent
problem given as the solution of the following equation:
\begin{equation}
  \label{eq:34}
  k_{\text{ET}} = \omega_e(k_{\text{ET}})
\exp\left[ -\langle \Delta E(k_{\text{ET}})\rangle^2 /2 C_X(0,k_{\text{ET}}) \right]
\end{equation}
Here, the rate-dependent energy gap can be re-written based on eq
\ref{eq:17} as
\begin{equation}
  \label{eq:36}
  \langle \Delta E(k_{\text{ET}}) \rangle = \Delta E^{\text{gas}} + \Delta E^{\text{ind}}+ 
  \Delta E^{\text{C}}_{\text{inh}} + f_{\text{ne}}^{\text{C}}(k_{\text{ET}}) \Delta E^{\text{C}}_r
\end{equation}
where, based on our simulations discussed below, we assume that the
induction component of the shift does not involve slow relaxation and
only Coulomb solvation gets cut off by breaking ergodicity.
Accordingly, the Gaussian width in eq \ref{eq:34} takes the form
\begin{equation}
  \label{eq:24}
  C_X(0,k_{\text{ET}}) = C^{\text{ind}}(0) + 2 k_{\text{B}}T \lambda^{\text{C}}(k_{\text{ET}})
\end{equation}
where
\begin{equation}
  \label{eq:37}
  \lambda^{\text{C}}(k_{\text{ET}}) = f_{\text{ne}}^{\lambda}(k_{\text{ET}}) \lambda^{\text{C}}
\end{equation}
In eqs \ref{eq:36} and \ref{eq:37}, we have introduced the parameters
of non-ergodicity of nuclear fluctuations contributing to the vertical
energy gap, $f_{\text{ne}}^{\text{C}}$, and to the reorganization
energy, $f^{\lambda}_{\text{ne}}$. The parameter $f^{\lambda}_{\text{ne}}$ can be
readily calculated from eqs \ref{eq:15}, \ref{eq:50}, and \ref{eq:18}:
\begin{equation}
  \label{eq:38}
  f_{\text{ne}}^{\lambda} =  \lambda_G/\lambda^{\text{C}} + (2/ \pi) \sum_{i=1,2} (\lambda_i/\lambda^{\text{C}}) 
                \cot^{-1} \left( k_{\text{ET}} \tau_i  \right) 
\end{equation}

The procedure outlined above can be used to construct the free energy
surfaces of electron transfer.  The widely accepted definition of the
free energy surfaces for electron transfer follows the general
procedure of defining the Landau functional\cite{Landau5} in which the
hypersurface $X=\Delta E$ generates the incomplete partition function
\begin{equation}
  \label{eq:29}
  e^{-\beta G(X)} \propto \int \delta\left(\Delta E -X \right) e^{-\beta H} d\Gamma  
\end{equation}
In this expression, $\Delta E$ depends on all nuclear modes $Q_n$,
$n=1,\dots, M$ in the system. In addition, $H$ is the system
Hamiltonian in the initial state of the electron-transfer system and
$d\Gamma$ is the element of phase space.

In applications to processes happening on short time scales, one needs
to generalize eq \ref{eq:29} to exclude a subset of frequencies not
contributing to the process:
\begin{equation}
  \label{eq:35}
   e^{-\beta G(k_{\text{ET}},X)} \propto \int  \delta\left(\Delta E -X \right) e^{-\beta H}
   \prod_{n,\omega<k_{\text{ET}}} \delta[Q_n(\omega)] d\Gamma
\end{equation}
In this equation, the product of delta functions eliminates
the low-frequency modes from the partition function.

\subsection{Polarizability of the special pair}
\label{sec:2-4}
There is a significant body of
experimental\cite{Lockhart:88,Haran:96,Wynne:96,Arnett:99} and
computational\cite{Larsson:90,Thompson:91,Lathrop:94,Zhou:97,Chang:01,Renger:04} evidence of a
strong mixing of covalent, (P$_L$-P$_M)^*$, and charge-transfer,
(P$_M^+$-P$_L^-)^*$, states within the photoexcited special pair, where P$_M$
and P$_L$ are the M and L subunits of the special pair (Figures
\ref{fig:2} and \ref{fig:3}).  Although the average amount of charge
transfer between two subunits is small,\cite{Thompson:91} about 0.1 of
the electronic charge in the gas phase and 0.2 in the reaction center, the
fluctuations of the electrostatic potential of the protein/water
solvent create significant fluctuations of the extent of charge
transfer.  Correspondingly, the fluctuating population of the
charge-transfer state (P$_M^+$-P$_L^-)^*$ creates fluctuating charges $\Delta
z_j=n_{\text{CT}} \Delta Z_j$ at the atomic sites of the special pair ($j$
runs over the atoms of P). In this representation, $\Delta Z_j$ are the
difference of atomic charges of the special pair between the ionized
excited state (P$_M^+$-P$_L^-)^*$ and the covalent state (P$_L$-P$_M)^*$,
and $n_{\text{CT}}$ is the population of the charge-transfer state.
Physically, this redistribution of charge in response to an external
electrostatic field implies that the special pair is polarizable with
the instantaneous induced dipole moment equal to
$\mathbf{p}_{\text{CT}}=n_{\text{CT}}\Delta\bm{\mu}_{\text{CT}}$. Here, $\Delta
\bm{\mu}_{\text{CT}}$ is the dipole moment between the ionized and
neutral states of the special pair. The importance of the induced
dipole moment for electron transfer is that the instantaneous electron
transfer dipole becomes modified from the dipole $\bm{\mu}_{\text{ET}}$
created by the set of permanent charges $\Delta q_k$ ($k$ runs over all
atoms of the cofactors involved in electron transfer) to a new
fluctuating dipole moment $\bm{\mu}_{\text{ET}} + \mathbf{p}_{\text{CT}}$.
Since the solvent reorganization energy is proportional to the average
squared dipole moment
\begin{equation}
  \label{eq:30}
  \lambda^{\text{C}} \propto \left\langle \left(\bm{\mu}_{\text{ET}} + \mathbf{p}_{\text{CT}}\right)^2\right\rangle  
\end{equation}
the appearance of the induced dipole can potentially modify the
energetics of electron transfer.\cite{DMjpca:99,DMjacs:03,DMjpcb2:06}

In order to model the effect of polarizability of the special pair on
the statistics of the donor-acceptor energy gap, we have adopted the
following simulation algorithm. The charges $z_j$ of the primary pair
are re-calculated at each fifth MD step according to the equation
\begin{equation}
  \label{eq:31}
  z_j = z_j^P + n_{\text{CT}} \Delta Z_j  
\end{equation}
where $z_j^P$ are the charges of two decoupled bacteriochlorophylls
obtained from our DFT calculations (Appendix \ref{secB} and supporting
information). The extent of charge delocalization $n_{\text{CT}}$ is
calculated by diagonalizing, at each fifth step of the MD trajectory,
the two-state quantum Hamiltonian characterized by the electronic
coupling $J$ and the instantaneous energy gap between two states
\begin{equation}
  \label{eq:32}
  \Delta \epsilon = \Delta \epsilon^{\text{gas}} + \Delta \epsilon^{\text{ind}} + (1/2) \sum_j \Delta Z_j \phi_j 
\end{equation}
Here, $\Delta \epsilon^{\text{gas}}$ is the gas-phase energy separation between
the neutral and ionized states of P and $\phi_j$ is the electrostatic
potential of the surrounding protein/water solvent at the position of
atomic charge $j$. $ \Delta \epsilon^{\text{ind}}$ in eq \ref{eq:32} is the
induction shift of the energy gap and the parameters $ \Delta
\epsilon^{\text{gas}} $ and $J$ are tabulated in Appendix \ref{secB}.

\subsection{Polarizable special pair and free energy surfaces of
  electron transfer}
\label{sec:2-5}
The description of Coulomb solvation presented in sec \ref{sec:2-1}
change significantly when the special pair is
polarizable.\cite{DMjpca:99,DMjacs:03} The main modification here is
that the atomic charges and hence the electric field of the cofactors
become a function of the solvent polarization $\mathbf{P}$ through the
extent of charge delocalization $n_{\text{CT}}$. The electric field
$\mathbf{E}_{01}$ changes from the value commonly calculated from the
vacuum charge distribution to $\mathbf{E}_{01}[\mathbf{P}]$:
\begin{equation}
  \label{eq:28}
  \mathbf{E}_{01} \to \mathbf{E}_{01}[\mathbf{P}] 
\end{equation}
A general solution for the free energies of electron transfer in this
case has not been found so far, although an analytical theory can be
formulated in the case of dipole solvation.\cite{DMjpca:01}
Alternatively, the field $\mathbf{E}_{01}[\mathbf{P}]$ can be linearly
expanded in the solvent polarization $\mathbf{P}$ around its
equilibrium value
\begin{equation}
  \label{eq:39}
  \mathbf{E}_{01}[\mathbf{P}] = \mathbf{E}_{0}[\mathbf{P}_{\text{eq}}+\mathbf{P}_{\text{eq}}'] 
+ \mathbf{F} \cdot\delta \mathbf{P}  
\end{equation}
where $\mathbf{F}$ is a 2-rank tensor and
$\delta\mathbf{P}=\mathbf{P}-\mathbf{P}_{\text{eq}} -
\mathbf{P}_{\text{eq}}'$.

When the form of the field given by eq \ref{eq:39} is substituted into
the standard Hamiltonian\cite{MarcusSutin} of the solute linearly
coupled to the Gaussian field $\mathbf{P}$, one gets
\begin{equation}
  \label{eq:40}
 H = - \mathbf{E}_{01}[\mathbf{P}_{\text{eq}}+\mathbf{P}_{\text{eq}}']*\mathbf{P} + 
               (1/2) \delta\mathbf{P}*\bm{\chi}_{\text{mod}}^{-1}*\delta\mathbf{P}
\end{equation}
where $\bm{\chi}_{\text{mod}}^{-1} = \bm{\chi}^{-1}- 2\mathbf{F}$ is the
new, modified linear response function of the Gaussian polarization
field renormalized by the solute polarizability. Since the
polarizability tensor $\mathbf{F}$ is generally different in the
initial and final electronic states, the donor-acceptor energy gap
becomes a bilinear function of the Gaussian field $\mathbf{P}$ in
contrast to the linear function used to derive the Marcus parabolas.
The main consequence of that change is that the statistics of energy
gap fluctuations become non-Gaussian.  The free energy surface loses
its parabolic shape predicted by Marcus theory and can instead be
represented by the analytical results of the Q-model:\cite{DMjcp:00}
\begin{equation}
  \label{eq:45}
  G(X) = \alpha \left(\sqrt{\left| \langle \Delta E \rangle - \alpha \lambda^{\text{C}} - X\right|}-\sqrt{\alpha \lambda^{\text{C}} } \right)^2 
\end{equation}
Here, $\alpha>0$ is the non-parabolicity parameter describing the deviation
of the free energy surface from the parabolic shape. The limit $\alpha \to \infty$
recovers the Marcus barrier thermodynamics.

\begin{table*}
  \centering
  \caption{{\label{tab:1}}Components of the average energy gap for primary charge separation (all energies are in eV).}
  \begin{tabular}{ccccccccc}
\hline 
Protocol &  $T$/K & $\Delta E^C_w$ & $\Delta E^C_{\text{prot}}$ & $\Delta E^C$ & $\Delta E^{\text{ind}}_w$ & $\Delta E_{\text{prot}}^{\text{ind}}$ & $\Delta E^{\text{ind}}$ & $\Delta E_s$ \\
\hline
S1 &  77 &  0.170  & -0.500  & -0.330 & -0.058  & -1.234 & -1.292 & -1.623 \\
   &  200 & 0.144  & -0.728  & -0.584 & -0.057  & -1.199 & -1.256 & -1.840 \\
   &  250 & 0.382  & -0.749  & -0.367 & -0.054  & -1.183 & -1.237 & -1.604 \\
   &  300\footnotemark[1] & 0.205  & -0.678  & -0.473 & -0.055  & -1.164 & -1.219 & -1.691 \\
   &  300\footnotemark[2] & -0.365 & -1.278 & -1.643 & -0.082 & -1.071 & -1.153 & -2.796 \\
   &  350 & 0.440  & -0.856  & -0.416 & -0.050  & -1.094 & -1.144 & -1.559 \\
   &  400 & 0.093  & -0.330  & -0.237 & -0.075  & -0.853 & -0.928 & -1.164 \\
S2 &  250 & 0.399  & -0.502  & -0.103 &  -0.048 & -1.042 & -1.090 & -1.193 \\
   &  275 & 0.404  & -0.535  & -0.131 &  -0.051 & -1.065 & -1.116 & -1.247 \\
   &  300\footnotemark[3] & 0.310  & -0.632  & -0.323 &  -0.052 & -1.076 & -1.128 & -1.451 \\
   &  325\footnotemark[3]  & 0.310  & -0.168  &  0.141 &  -0.020 & -0.836 & -0.857 & -0.716 \\
   &  350\footnotemark[3]  & 0.347  & -0.594  & -0.247 &  -0.053 & -0.987 & -1.041 & -1.287 \\
\hline
  \end{tabular}
\footnotetext[1]{Obtained from 10 ns long MD trajectories, the unmarked
  data refer to 5 ns of simulations.}
\footnotetext[2]{Data for the final charge transfer state
  P$^+$--B$_L^-$, 5 ns trajectory. }
\footnotetext[3]{Obtained from 15 ns long MD trajectories.}
\end{table*}

\section{Results}
\label{sec:3}
We believe that this paper reports the most extensive MD simulations
on the bacterial reaction center following previous simulation efforts
in this
field.\cite{Warshel:89,Treutlein:92,Marchi:93,Gehlen:94,Parson:98,Ceccarelli1:03,Sterpone:03}
The overall length of 118 ns of MD trajectories, of which 100 ns were
used for the production analysis, required 39.8 CPU years. All
simulations were done in parallel using 128 CPUs of ASU's HPC
facility. The analysis of the simulations was performed by a parallel
code developed for this project that reads directly binary AMBER
files. The analysis was run in parallel on 10 Opteron CPUs and
required overall 4.8 years of CPU time.

Two sets of simulations have been done. The first set, which we will
label S1, was performed at six different temperatures.  It employed
the standard protocol of MD force fields with fixed atomic charges.
The equilibrium MD trajectories were used to calculate the statistics
of the donor-acceptor energy gap and the Stokes shift correlation
functions. In this calculation, in addition to Coulomb interactions,
induction solute-solvent interactions were computed. The atomic
polarizabilities were taken from a modified Thole
parametrization.\cite{vanDuijnen:98} The induction potential was not a
part of the simulation algorithm, thus assuming that the exploration
of the phase space of the nuclear motions can be accomplished with the
standard force fields. Since these force fields effectively
incorporate polarizability in terms of permanent charges, in order to
avoid double counting, the charges of the solvent (protein and water)
were multiplied by 0.89 in analyzing the data, following the
convention adopted in the literature.\cite{Sterpone:03}

Six trajectories of S1 protocol were produced for the initial state of
the reaction complex, (P--B$_L)^*$, at different temperatures. The
atomic partial charges calculated by us at the DFT level (Appendix
\ref{secB}) were supplemented by the force-field parameters of
bacteriochlorophyll developed by Marchi and
co-workers.\cite{MarchiParms02} The atomic charges of the ground-state
bacteriochlorophyll were used for the exited state of the primary pair
assuming that photoexcitation does not greatly alter the charge
distribution.\cite{Parson:98} One simulation trajectory at 300 K was
produced for the charge-separated state P$^+$--B$_L^-$ corresponding
to the first hop of the electron in the sequential mechanism. For this
simulation, the positive charge of P$^+$ was distributed among the two
cofactors of the special pair as described in Appendix \ref{secB} and
the charge distribution of the bacteriochlorophyll anion was calculated at
the DFT level (supporting information).

The second set of simulations, labeled as S2, required changing the
standard MD protocol (see Appendix \ref{secB}). In these simulations,
quantum polarizability of the special pair was accounted for by
diagonalizing the 2$\times$2 Hamiltonian matrix of the charge-transfer
state between the two parts of P at each fifth step of the MD
trajectory, a procedure known in the literature as the empirical
valence bond approach.\cite{Aqvist:93,Schmitt:99} The Hamiltonian
diagonalization allows one to calculate the extent of charge transfer
between two bacteriochlorophylls in P and dynamically adjust charges of
the special pair. This simulation protocol thus incorporates an
extremely high polarizability of P$^*$ revealed by Stark spectroscopy
measurements of Boxer and co-workers.\cite{Lockhart:88,Middendorf:93}

\begin{table*}
  \caption{{\label{tab:2}}Reorganization energies calculated from fluctuations of 
    the energy gap (eq \ref{eq:46}). All energies are in eV. }
\begin{tabular}{ccccccccccc}
\hline
Protocol & T/K  &  $\lambda^{\text{ind}}_{\text{w}}$ &
$\lambda^{\text{ind}}_{\text{prot}}$  & $\lambda^{\text{ind}}$ 
         &      $\lambda^{\text{C}}_{\text{w}}$  &
         $\lambda^{\text{C}}_{\text{prot}}$   & $\lambda^{\text{C}}$
         &  $\lambda_{\text{w}}$ & $\lambda_{\text{prot}}$  & $\lambda_{s}$\footnotemark[1]\\ 
\hline
S1 & 77   &  0.019  &  0.065  & 0.070 & 0.187  &  0.168  & 0.351 & 0.191 & 0.245 & 0.421 \\ 
   & 200  &  0.001  &  0.057  & 0.062 & 0.756  &  0.182  & 0.845 & 0.756 & 0.251 & 0.903 \\
   & 250  &  0.016  &  0.076  & 0.081 & 1.634  &  0.341  & 1.938 & 1.639 & 0.419 & 1.955 \\ 
   & 300\footnotemark[2]  &  0.047   &  0.112  & 0.119 & 1.136  &  0.375  & 1.564 & 1.124 & 0.466 & 1.598 \\  
   & 300\footnotemark[3]  &  0.047  &  0.146  & 0.149 & 1.393  &  0.441  & 1.542 & 1.379 & 0.593 & 1.692 \\
   & 350  &  0.110  &  0.187  & 0.191 & 0.948  &  0.644  & 1.508 & 0.944 & 0.853 & 1.508 \\ 
   & 400  &  0.139  &  0.249  & 0.275 & 0.767  &  0.567  & 1.010 & 0.866 & 0.797 & 1.335 \\
S2   & 300  &  0.481  &  0.682  & 0.697 & 1.439  &  0.735  & 1.839 & 1.454 & 1.385 & 2.513 \\
\hline 
\end{tabular}
\footnotetext[1]{$\lambda^{\text{ind}}+\lambda^{\text{C}}$ deviates slightly from $\lambda_s$ because 
of numerical uncertainties of averaging.}
\footnotetext[2]{Obtained from 10 ns long MD trajectories, the unmarked
  data refer to 5 ns of simulations.}
\footnotetext[3]{5 ns data for the final charge transfer state
  P$^+$--B$_L^-$.}
\end{table*}

\subsection{Energetics}
\label{sec:3-1}
Two energy parameters are of main importance within the Gaussian
picture of electron transfer activation (Marcus model). These are the
average donor-acceptor energy gap and the energy gap variance (eq
\ref{eq:2}). These parameters, obtained from MD simulations at
different temperatures, are listed in Tables \ref{tab:1} and
\ref{tab:2}. The complete set of first cumulants from both S1 and S2
simulations is reported in Table \ref{tab:1}. The S2 entry in Table
\ref{tab:2} is limited to 300 K since the second cumulants at other
temperatures did not converge on the time-scale of the simulation
trajectories.

Since we are dealing here with a heterogeneous solvent composed of
a protein matrix and aqueous environment, the separation of these
two first cumulants of the energy gap into the water and protein
contributions provides mechanistic insights into the factors
influencing electron transfer activation. In addition, we split the
relevant energies into contributions from non-polar and Coulomb
interactions.  Finally, the introduction of polarizability (charge
fluctuations) of the special pair shifts relative weights of each
component in the activation barrier and, more importantly, results in
significant deviations from the Gaussian picture of Marcus
parabolas.
 
\begin{figure}
  \centering
  \includegraphics*[width=7cm]{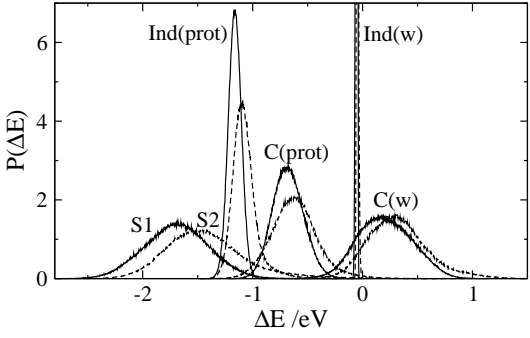}
  \caption{Normalized distributions of components of the
    donor-acceptor energy gap in non-polarizable (S1, solid lines) and
    polarizable (S2, dashed lines) simulation protocols ($T=300$
    K). Marked in the plot are the Coulomb interaction due to the protein
    (C(prot)) and water (C(w)) and the induction interaction
    (Ind(prot) for the protein and Ind(w) for water). S1 and S2 mark the
    distributions of the total energy gap for the non-polarizable (S1) and
    polarizable (S2) special pair.  }
  \label{fig:4}
\end{figure}

Figure \ref{fig:4} reports the distribution of Coulomb and induction
components of the energy gap from simulations of both the non-polarizable
and polarizable special pair. The Coulomb interactions have Gaussian
statistics where the width is consistent with the reorganization
energies listed in Table \ref{tab:2}. The average shifts arising from
water and the protein have opposite signs. Therefore, the polarization of
water by the protein matrix contributes to the destabilizing of the
charge-transfer state, as was also observed by Parson \textit{et
  al}.\cite{Parson:98} On the contrary, the protein matrix makes the
dominant contribution into stabilizing the charge-separated state.
The induction shift of the average energy gap, arising primarily from
the protein matrix (Ind(prot) in Figure \ref{fig:4}), is about twice
larger than the Coulomb shift which largely cancels out between
its protein and water contributions (Table \ref{tab:1}). On the
contrary, the width of the distribution of induction energies is small
relative to the Coulomb interactions for non-polarizable (S1)
simulations (in accord with assessment of analytical
theories\cite{DMjcp:95}), but grows significantly for the polarizable
(S2) simulation protocol (Table \ref{tab:2}).

The splitting of the total self-correlation function $C_X(0)$ into the
individual protein (subscript ``prot'') and water (subscript ``w'')
components requires an estimate of the cross-correlation term
$\lambda_{\text{w,prot}}$ between the water and protein interaction
potentials:
\begin{equation}
  \label{eq:52}
  \lambda_s = \lambda_{\text{prot}} + \lambda_{\text{w}} + \lambda_{\text{prot,w}}  
\end{equation}
This latter part turns out to be significantly smaller than the
individual protein and water components, as can be inferred from the
last three columns in Table \ref{tab:2} by comparing the total solvent
reorganization energy $\lambda_s$ with the sum of the two components, $
\lambda_{\text{prot}} + \lambda_{\text{w}}$.

Notwithstanding such little attention paid in the electron-transfer
literature to non-polar interactions, the induction shift is the main
part of the solvent effect on the average energy gap of charge
separation. Its value can be estimated from some simple arguments. If
one assumes that atomic polarizabilities are distributed with a
constant density around the donor and acceptor, one arrives at a simple
expression\cite{DMjcp2:05}
\begin{equation}
  \label{eq:26}
  \Delta E^{\text{ind}}_{\text{prot}} = - 3e^2 \frac{n_{\text{prot}}^2
    -1}{n_{\text{prot}}^2 + 2} \left(\frac{1}{2R_D}+\frac{1}{2R_A} - \frac{1}{R_{DA}}\right)
\end{equation}
Here, $R_D$ and $R_A$ are the radii of the donor and acceptor and
$R_{DA}$ is the distance between them. In addition, $n_{\text{prot}}$
is the refractive index of the protein matrix and $e$ is the
elementary charge. For the average refractive index of the reaction
center\cite{protRefInd} $n_{\text{prot}}=1.473$ and the radius of the
bacteriochlorophyll unit $R_D=R_A=5.6$ \AA{} obtained from its vdW volume
one gets $ \Delta E^{\text{ind}}_{\text{prot}}=-1.09$ eV at the crystallographic distance
$R_{DA}=11.3$ \AA. This number compares favorably with the induction
shift of $\Delta E^{\text{ind}}_{\text{prot}} = -1.16$ eV from MD simulations at $T=300$ K
(Table \ref{tab:1}, S1 protocol).

The positive slope of the induction shift of the average energy gap is
caused by the temperature expansion of the protein. Based on the data
shown in Table \ref{tab:1} for S1 simulation protocol, the logarithmic
derivative of the induction shift with temperature, $d\ln \Delta
E^{\text{ind}} /dT$, is within the limits $(4-8)\times 10^{-4}$
K$^{-1}$. According to eq \ref{eq:26} this derivative should be equal
to thermal expansivity of the protein (Clausius-Mossotti equation).
Indeed, the logarithmic slope of the induction shift agrees reasonably
well with the reported\cite{Sasisanker:04} expansion coefficients of
proteins of the order of $8 \times 10^{-4}$ K$^{-1}$.
 
Several of the MD simulation results reported here turned out to be
quite surprising. Among the unexpected findings are quite large values
of the solvent (protein and water) reorganization energies,
contrasting the commonly low values (ca.\ 0.1--0.2 eV) circulating in
the
literature.\cite{Treutlein:92,Bixon:95,Parson:98,Haffa:02,Sterpone:03}
In particular, water is far from being screened by the
protein\cite{Sulpizi:07} making the main portion of the energy gap
variance in the S1 protocol, and being surpassed by the protein in the
S2 protocol. In fact, the values of water reorganization energy found
here are more typical of small redox couples in aqueous
solution\cite{Newton:99} than of the often anticipated hydrophobic
screening by the protein matrix.

\begin{figure}
  \centering
  \includegraphics*[width=7cm]{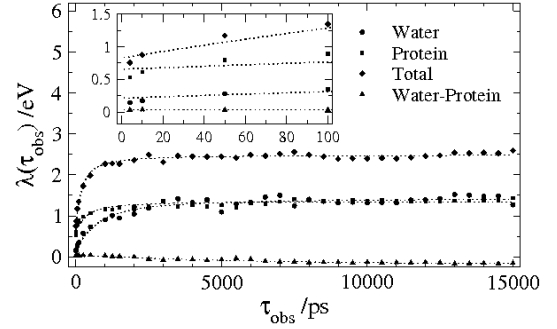}
  \caption{Components of the solvent reorganization energy from water
    and protein from MD simulations vs the observation time
    $\tau_{\text{obs}}$. Points refer to MD data (S2 protocol, 300 K) and
    the dotted lines indicate the fits to eq \ref{eq:47}. The inset
    shows the initial portion of the plot; the relaxation times used
    to fit the MD results to eq \ref{eq:47} are: $\tau_R = 218$ ps
    (total), $\tau_R = 390$ ps (protein), and $\tau_R = 764$ ps (water).  }
  \label{fig:5}
\end{figure}

What is different in our simulations compared to previously reported
simulation data\cite{Treutlein:92,Parson:98,Sterpone:03} is the length
of the simulation trajectory which has allowed us to push the numbers
for the reorganization energies closer to their thermodynamic limit.
Indeed, as is seen in Figure \ref{fig:5}, the reorganization energy as
a function of the length of the simulation trajectory (observation
time $\tau_{\text{obs}}$) levels off by the end of the 5--10 ns production
run. However, this long-time reorganization energy is not relevant for
the short-time charge-separation dynamics since a significant subset
of nuclear modes gets dynamically arrested on the picosecond
time-scale at which the reorganization energy as a function of
$\tau_{\text{obs}}$ starts to sharply decline (Figure \ref{fig:5}). The
dependence of the total reorganization energy and its components on
the observation window can be fitted to a one component Debye equation
(\textit{cf}.\ to eq \ref{eq:38})
\begin{equation}
  \label{eq:47}
   \lambda(\tau_{\text{obs}}) \propto \cot^{-1}(\tau_R / \tau_{\text{obs}}) 
\end{equation}
with the effective relaxation time $\tau_R$. The fits shown by dotted
lines in Figure \ref{fig:5} indicate that the system starts to lose
ergodicity on the time-scale of several hundred picoseconds.

The data in Figure \ref{fig:5} have been generated according to the
following procedure.  First, a trajectory of individual
protein/solvent vertical energies is created from the sum of their
respective Coulomb and induction components.  Second, a smaller
trajectory window of length $\tau_{\text{obs}}$ is cut from the full MD
trajectory.  Third, the energy gap variance is calculated on this
smaller observation window which is then moved along the entire
trajectory. Each time the window is shifted, the variance is
calculated with the average energy gap set to its average on that
particular window.  The individual variances are then averaged among
the results from each sliding window, and the average reorganization
energy is reported as the $\lambda(\tau_{\text{obs}}$) in Figure \ref{fig:5}.

\begin{figure}
  \centering
  \includegraphics*[width=7cm]{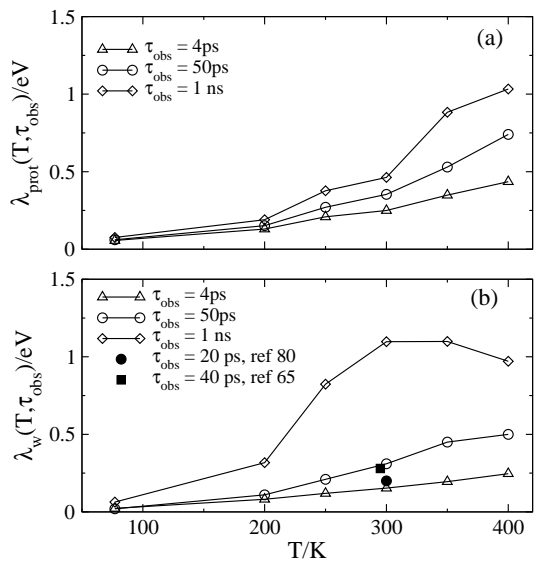}
  \caption{Protein (a) and solvent (b) reorganization energies from MD
    simulations as functions of the observation time $\tau_{\text{obs}}$
    and temperature. The closed square indicates the result of ref
    \onlinecite{Parson:98} obtained on a 40 ps observation window and
    the closed circle refers to the 20 ps window used in ref
    \onlinecite{Treutlein:92}.  }
  \label{fig:6}
\end{figure}

Reorganization energies calculated from this algorithm are plotted vs
temperature in Figure \ref{fig:6}. As in Figure \ref{fig:5},
shortening the observation window lowers the reorganization energy. On
the 4 ps observation window, most of the multi-exponential Stokes
shift relaxation is dynamically arrested (see below) and only
ballistic Gaussian relaxation from the Coulomb interactions and the
modulation of induction interactions by density fluctuations
contribute to the reorganization energy. For this short observation
time, the reorganization energy falls in the range of values commonly
reported from fitting the experimental reaction
rates.\cite{Bixon:95,Haffa:02} In particular, our results from 50 ps
observation window are consistent with the previous report by Parson
\textit{et al}.\cite{Parson:98} using 40 ps of the simulation
trajectory (for \textit{Rp.\ viridis}), while the result from 20 ps
simulations from Treutlein \textit{et al.}\cite{Treutlein:92} is
slightly below that value (closed points in Figure \ref{fig:6}b). We
do not expect a close agreement here since our algorithm of sliding
window generally gives rise to higher reorganization energies than a
single observation.

The reorganization energy from the protein is an increasing function of
temperature for all observation windows (Figure \ref{fig:6}a). On the
contrary, for water reorganization, the negative temperature slope
expected from equilibrium statistical mechanics\cite{DMjpca1:06} is
reverted by non-ergodicity to a positive one (Figure \ref{fig:6}b). This
effect is caused by a temperature-depending unfreezing of the nuclear
modes when relaxation becomes faster with increasing temperature. The
downward turnover of $\lambda(T)$ for the 1 ns observation window (upper
curve in Figure \ref{fig:6}b) marks the return of the system to
equilibrium statistics with the negative slope of $\lambda(T)$ also seen in
our previous simulations of a small solute in SPC/E
water.\cite{DMjpcb1:06}

\begin{figure}
  \centering
  \includegraphics*[width=7cm]{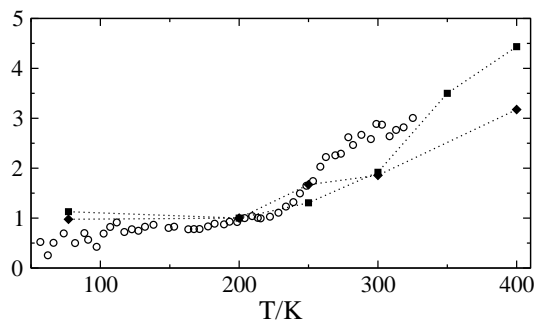}
  \caption{Induction (squares) and total protein (diamonds)
    reorganization energies from the present MD simulations and
    experimental mean square displacements of hydrogens of
    bacteriorhodopsin scaled with the inverse temperature, $\langle (\delta x)^2\rangle
    /T $ (circles). The experimental data were obtained by neutron
    scattering.\cite{Ferrand:93} All parameters have been normalized
    to their corresponding values at 200 K.  }
  \label{fig:7}
\end{figure}

The opposite temperature dependence of the protein and water
reorganization energies at long observation windows points to a
distinctly different character of the corresponding nuclear modes.
While water molecules alter the donor-acceptor energy gap mostly by
librational/rotational motions typical of polar liquids, the protein
nuclear modes are predominantly vibrational.  The temperature
dependence of $\lambda_{\text{prot}}$ seems to correlate well with the
temperature dependence of atomic displacements of the protein matrix
as is illustrated in Figure \ref{fig:7} where we show the better
converged induction reorganization energy along with the total protein
reorganization energy. The temperature change of these reorganization
energies is compared with mean square displacements of hydrogens in
bacteriorhodopsin obtained from inelastic neutron
scattering.\cite{Ferrand:93}

It is by now well established that protein vibrations start to deviate
from the straight line of harmonic motions at the transition
temperature of about $T_r\simeq 200-220$ K marking the rise of anharmonic
motions (including side-chain
rotations).\cite{Ferrand:93,Bizzarri:04,Fenimore:04} Therefore, the
mean-square displacement scaled with inverse temperature, $\langle(\delta
x)^2\rangle/T$, is a flat function at low temperatures starting to rise
above the transition temperature $T_r$.  The same trend is seen for
the total protein reorganization energy and its induction component,
which both turn to a sharp increase at about the same temperature.
This comparison implies that the relatively large values of protein
reorganization energy obtained in our simulations at room temperature
can be traced back to highly anharmonic motions of the protein matrix.

\subsection{Polarizable special pair}
\begin{figure}
  \centering
  \includegraphics*[width=7cm]{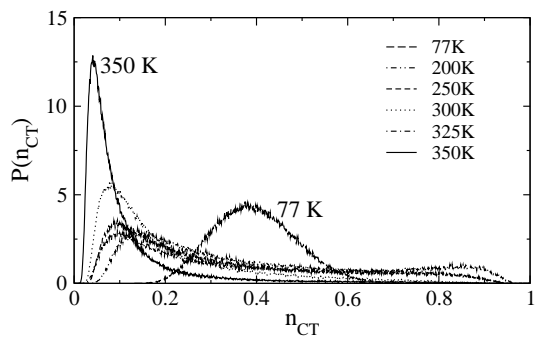}
  \caption{Distribution of the population number of the ionized state
    of the special pair along the simulation trajectory at different
    temperatures. The length of the simulation trajectory varies from
    5 ns at 77 K to 15 ns at 300, 325, and 350 K. }
  \label{fig:8}
\end{figure}

We need to emphasize here that our modeling of the polarizability of
the special pair carries qualitative significance only. In addition to
the obvious limitations of the two-state model, the modeling of the
temperature dependence of the special pair polarizability is not
adequate.  In our current simulations, the temperature dependence of
the average population of the ionized, charge-transfer state of the
special pair, $n_{\text{CT}}(T)$, originates from the temperature
dependence of the diabatic diagonal energy gap (eq \ref{eq:32}). This
component of the two-state Hamiltonian increases with growing
temperature, in general agreement with the idea that a polar
environment should become effectively less polar with increasing
temperature. Therefore, as is illustrated in Figure \ref{fig:8}, the
special pair becomes effectively more localized at higher temperatures
because the average energy splitting between the two state grows with
increasing temperature. The broad distribution of $n_{\text{CT}}$ is a
signature of the strong vibronic coupling of the charge-transfer
state.\cite{Zhou:97} What effectively happens due to strong
temperature dependence of the average population is that the
polarizability of the special pair is about 400 \AA$^3$ at $T=300$ K
increasing up to 1800 \AA$^3$ at 77 K.  Given the experimental
temperature variation of the absorption band of the special
pair\cite{Kirmaier:88} and the results of Stark spectroscopy at 77
K,\cite{Middendorf:93} the former values appears to be more realistic
than the latter.

The increase of localization of the special pair in its neutral
(P$_M$-P$_L)^*$ state results in a blue shift of the absorption
spectrum in general agreement with
experiment.\cite{Kirmaier:88,Huber:98} However, the slope of this
temperature dependence derived from the data shown in Figure
\ref{fig:8} appears to be too large. In vibronic models of the
temperature effect on the special pair absorption this effect is
modeled by temperature-dependent population of vibronic modes coupled
to the dimer charge-transfer state.\cite{Renger:04,Renger:07} This
implies the temperature shift of the diabatic diagonal energy
gap. Since this property is determined by the protein/water
electrostatic potential in our simulations, a possible way to off-set
a too strong temperature dependence of absorption frequency is to
introduce a temperature-dependent off-diagonal coupling $J$ (Appendix
\ref{secB}).\cite{Chang:01} Low frequency vibrations of the special
dimer in the 90--160 cm$^{-1}$ region\cite{Vos:99} might contribute to
that temperature dependence.  It seems that the model needs to be modified
to reproduce the temperature variation of the absorption spectrum of
the special pair. The current simulations in S2 protocol are therefore
not capable to properly address the issue of the temperature
dependence of the rate. However, we still believe that our results
provide valuable insights into how the parameters of the model change
once the polarizability is turned on. We therefore report the results
of simulations here with the warning that the parameter magnitudes
might be modified with the refinement of the model. We will also limit
our analysis of the free energy surfaces of electron transfer to 300 K
at which the polarizability seems to be more realistic. What this
value at room temperature should be is not entirely clear since the
Stark data were reported at 77 K\cite{Lockhart:88} (see Appendix
\ref{secB}).

\subsection{Free energy surfaces}
\label{sec:3-2}
A general solution for the non-ergodic free energy surface defined by eq
\ref{eq:35} is still missing. The current calculations and analysis of
MD data are therefore limited to the phenomenological approach
outlined in sec \ref{sec:2-3} where a step-wise frequency filter was
introduced into the frequency linear response functions. Computer
simulations and comparison to optical experiments in glass-forming
liquids support this approach\cite{DMacc:07} and one therefore can ask
what would be the free energy surface $G(k_{\text{ET}},X)$ on the
time-scale of primary charge separation
$\tau_{\text{ET}}=k_{\text{ET}}^{-1}$ compared to the thermodynamic
surface $G(X)$.

\begin{figure}
  \centering
  \includegraphics*[width=7cm]{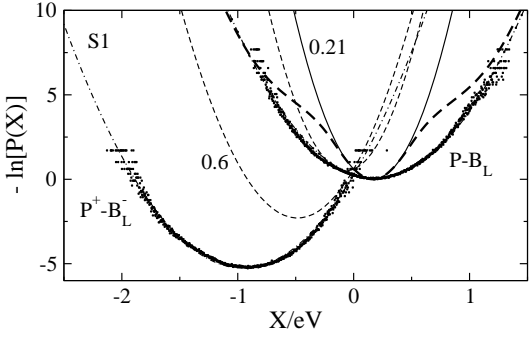}
  \caption{Free energy surfaces of primary charge separation obtained
    from MD simulations of reaction center with non-polarizable
    (constant charges, S1) special pair. The free energy surfaces $\beta
    G(X)=-\ln P(X)$ have been obtained from the normalized
    distributions of the total energy gap $X$ from MD simulations
    (points) in the initial charge-transfer state, P--B$_L$ (10 ns
    trajectory), and the final state, P$^+$--B$_L^-$ (5 ns
    trajectory).  The pair of curves marked with 0.21 are the
    non-ergodic free energy surfaces calculated by using
    $f_{\text{ne}}^{\lambda}=0.21$ value of the non-ergodicity parameter
    following from the fit of the theoretical rate to experiment
    ($T=300$ K). The vertical separation of the initial and final free
    energy curves is $+97$ cm$^{-1}$. The vertical separation of
    $-450$ cm$^{-1}$ reported by Zinth and co-workers\cite{Schmidt:94}
    is obtained when $f_{\text{ne}}^{\lambda}=0.6$ (marked in the plot) is
    used in the calculation of the final free energy curve. The
    vertical separation between ``equilibrium'' curves is $-1100$
    cm$^{-1}$. The dash-dotted lines in the plot are fits to Marcus
    parabolas yielding equal reorganization energies $\lambda_s\simeq 1.6$ eV
    consistent with direct calculations of second energy gap cumulants
    in Table \ref{tab:2}.  The bold dashed line indicates the free
    energy obtained by solving the self-consistent non-ergodic
    equation for the rate (eq \ref{eq:34}) by varying the average
    energy gap (see the text). The parameters are those used to
    calculate the charge-separation rate in Figure \ref{fig:14}. }
  \label{fig:9}
\end{figure}

\begin{figure}
  \centering
  \includegraphics*[width=7cm]{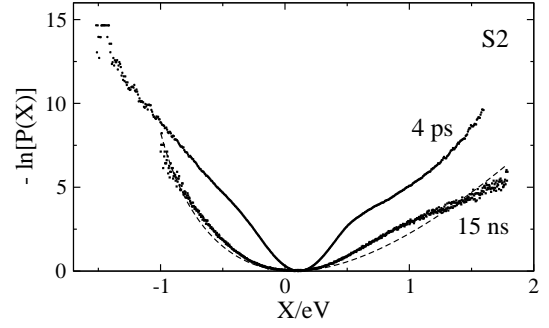}
  \caption{Free energy surfaces of primary charge separation obtained
    from MD simulations of the reaction center with polarizable
    (fluctuating charges, S2) special pair.  The upper curve is
    obtained from the simulation analysis with a 4 ps observation
    window, while the lower curve refers to the observation window of
    15 ns. The dashed line is the fit of the 15 ns simulation data to the analytical
    Q-model with the fitting parameters: $\langle\Delta E\rangle =0.07$ eV, $\lambda_s=2.81$
    eV, and $\alpha = 0.45$. }
  \label{fig:10}
\end{figure}

The thermodynamic free energy surface is of course not available to us
since sampling is always an issue with simulations. However, leveling
off of the reorganization energies on the 10--15 ns trajectory seen in
Figure \ref{fig:5} allows us to hope that, except for the slowest
modes responsible for the conformational mobility of the protein, the
phase space relevant to activating charge separation was adequately
sampled. The free energy surfaces for non-polarizable (S1) simulations
obtained from the 10 ns trajectory for the initial (P--B$_L)^*$ state
and from the 5 ns trajectory for the final (P$^+-$B$_L^-)^*$ state are
shown in Figure \ref{fig:9}. The results of polarizable (S2)
simulations are collected in Figure \ref{fig:10}. Our simulations
allow us to sample only the total interaction between the cofactors
and the protein/water solvent (eq \ref{eq:25}) and therefore the
gas-phase energy gap is missing from the overall energy gap $X$. This
component of the energy gap was obtained from fitting the calculated
rate constants at 300 K to the experimental data by Fleming \textit{et
  al}\cite{Fleming:88} and Wang \textit{et al}\cite{Wang:07} (see
below). The gas-phase gap obtained from the fit $\Delta E^{\text{gas}} =
1.86$ eV was used to horizontally shift $G(X)$ obtained from
simulations resulting in the average energy gap of $\langle \Delta E\rangle = 0.169$
eV. This number, which is equal to the energetic separation of the
free energy minimum from the point of activationless electron transfer,
is consistent with the experimental value of 0.127--0.147 eV (from
mutagenesis data) which separates the wild type reaction center from
the top of the Marcus inverted parabola.\cite{Haffa:02} Our result is
also close to $\langle \Delta E\rangle = 0.150$ eV reported by Wang \textit{et
  al}\cite{Wang:07} from fitting experimental data to the
diffusion-kinetic model (see below).

Long-trajectory simulations in S1 protocol produce Marcus parabolas
(dashed lines in Figure \ref{fig:9}) with the curvatures reproducing
reorganization energies listed in Table \ref{tab:2}.  Figure
\ref{fig:9} also shows the non-ergodic parabolas. Before explaining
the calculation of those, we first need to comment on the experimental
preparation of the initial state for charge separation. The initial
state for primary charge separation is prepared by photoexcitation of
the special pair which prior to that stays in the ground state for a
time long compared to any relaxation time in the system. The ground
state is thus characterized by the equilibrium polarization
$\mathbf{P}_{\text{eq}}'+\mathbf{P}_{\text{eq}}$ of which
$\mathbf{P}_{\text{eq}}'$ is the result of the inhomogeneous
protein/water environment and $\mathbf{P}_{\text{eq}}$ comes from
the polarization of the environment by the special pair.  When lifted to
the excited state by the absorbed photon, the special pair changes its
charge distribution and the polarization $\mathbf{P}_{\text{eq}}$ can
dynamically adjust to the new equilibrium polarization
$\mathbf{P}_{\text{eq}}^*$. We will assume that this change,
$\mathbf{P}_{\text{eq}}^* - \mathbf{P}_{\text{eq}}$ is insignificant
compared to $\mathbf{P}_{\text{eq}}'+\mathbf{P}_{\text{eq}}$ on the
reaction time-scale. We will therefore neglect the non-ergodicity
correction in the Coulomb component of the shift assuming $\Delta
E^{\text{C}}(k_{\text{ET}})=\Delta E^{\text{C}}$. This approximation
results in the following non-ergodic free energy surface
\begin{equation}
  \label{eq:53}
  G(k_{\text{ET}},X) = \frac{(X-\langle\Delta E\rangle )^2}{4 \lambda(k_{\text{ET}})}
\end{equation}
In this equation, 
\begin{equation}
  \label{eq:54}
  \lambda(k_{\text{ET}}) = \lambda^{\text{ind}} + f_{\text{ne}}^{\lambda}(k_{\text{ET}}) \lambda^{\text{C}}  
\end{equation}
is the non-ergodic reorganization energy affected by the dynamical
arrest of the Coulomb component of the solvent reorganization. The fit
of the rate constant to experiment (see below) results in $
f_{\text{ne}}^{\lambda} =0.21$, and the free energy surface obtained by
using this non-ergodicity parameter is shown by the solid line in
Figure \ref{fig:9}.

There is a significant difference between the way the initial and
final states for the first electron hop are created. The final state
is characterized by an instantaneously created dipole moment of the
charge-separated state and so the corresponding Stokes shift requires
non-ergodic correction with the following result for the final free
energy surface
\begin{equation}
  \label{eq:55}
  G'(X) = \frac{(X-\langle\Delta E\rangle + f_{\text{ne}}' \Delta
    X_{\text{st}})^2}{4\lambda' }
  + \Delta G_{\text{ne}}
\end{equation}
In this equation, $\Delta X_{\text{st}}$ is the total Stokes shift between
the minima of two parabolas achieved on long simulation trajectories.
The non-ergodicity parameter $f'_{\text{ne}}$ and the reorganization
energy $\lambda'$ depend on both the life-time of the charge-separated state
and the corresponding Stokes shift dynamics.  Finally, the vertical
energetic separation between the parabolas' minima $\Delta G_{\text{ne}}$
(not the reaction free energy) follows from the condition
$G(k_{\text{ET}},0)=G'(0)$ once all other parameters are known.

We currently do not have sufficient data to calculate the non-ergodic
parameters in eq \ref{eq:55} (which require, among other things, the
free energy surface corresponding to the electron located at H$_L$) and so
will limit our arguments to qualitative considerations only. The
equilibrium free energy surfaces obtained from long simulation
trajectories are vertically shifted by $\Delta G=-1100$ cm$^{-1}$. This
number is consistent with experimental data from recombination
rates\cite{Volk:98,Winter:03} which have put the lowest limit for $\Delta
G$ at $\simeq -2000$ cm$^{-1}$. This later value might be overestimated
since it was measured on the 100-$\mu$s lifetime of the triplet state of
the special pair. Delayed fluorescent measurements\cite{Ogrodnik:94}
sampling the system on the 20 ns time-scale and photovoltage
measurements at the 15 ns time-scale,\cite{Trissl:01} both comparable to
the length of simulations, show somewhat smaller gaps, $\Delta G_{\text{ne}} \simeq -1370$
cm$^{-1}$ and $-1180$ cm$^{-1}$, respectively. The latter data refer,
however, to the \textit{Rhodospirillum rubrum} reaction center.

We need an assignment of $f'_{\text{ne}}$ in eq \ref{eq:55} to produce
the non-ergodic free energy surface of the charge-separated state.  If
we use $f'_{\text{ne}}=f_{\text{ne}}^{\lambda}=0.21$ from the analysis of
the primary charge separation rate, we get essentially no vertical
shift of the two parabolas, $\Delta G_{\text{ne}}= 97$ cm$^{-1}$. The
vertical shift of $-450$ cm$^{-1}$ reported by Zinth and
co-workers\cite{Schmidt:94} is obtained when $f'_{\text{ne}} = 0.6$ is
used in eq \ref{eq:55}. This latter value of the vertical displacement
of parabolas minima, measured on the picosecond time-scale, compares
well with the estimate by Holzwarth and M{\"u}ller,\cite{Holzwarth:96}
$-331$ cm$^{-1}$, also done on the picosecond scale.  We will postpone
a more detailed analysis of the energetics of subsequent electron hops
to a future publication, while the current analysis is aimed to show
that overall our results do not contradict the key experimental
observations reported in the literature. We only note here in passing
that, similar to our previous simulations of hydrated
plastocyanin,\cite{DMjpcb1:08} our present simulations show a clear
separation between the Stokes shift $\Delta X_{\text{st}}$ and twice the
solvent reorganization energy, $2\lambda_s$ (also see ref
\onlinecite{Parson:98}). We will address this problem in more detail
elsewhere.\cite{DMjpcb3:08}

We need to caution here against a too literal understanding of the
non-ergodic free energy surfaces of electron transfer. Under ergodic
conditions, the free energy surface can be sampled by changing the
average energy gap by, for instance, optical spectroscopy. The result
can then be directly applied to the Frank-Condon factor of the
reaction yielding the reaction energy gap law. In the case of
non-ergodic reactions, this direct application of the free energy
surface obtained at a given observation window is prohibited since the
spectrum of fluctuations changes at each rate constant achieved by
horizontally sliding the free energy surface and thus sampling the
average gap. In order to illustrate that, we have plotted in Figure
\ref{fig:9} the free energy surface obtained by changing the average
energy gap in the self-consistent non-ergodic equation for the rate
constant (eq \ref{eq:34}). The result is a funnel-like surface, which
we also previously obtained in a study of ergodicity breaking in
liquid crystals.\cite{DMjpcb3:06} The such obtained curve transforms
from the narrow free energy surface at a high reaction rate to the
thermodynamic surface when the barrier for the reaction increases and
the rate slows down.

As is clear from the broader distribution of energy gaps for the
polarizable special pair (Figure \ref{fig:3}) and from Table
\ref{tab:2} where specific values of the reorganization energies are
listed, the free energy surfaces $G(X)$ are quite different for
a polarizable and non-polarizable special pair. As a matter of fact, not
only curvatures (reorganization energies) are different in two cases,
but also the shape of the free energy surface changes from a Marcus
parabola in the former case to a significantly asymmetric shape
in the latter (Figure \ref{fig:10}).  This result is consistent with
the predictions of the Q-model of electron transfer in polarizable
donor-acceptor complexes and, in fact, the simulated curve is well
fitted to eq \ref{eq:45} (dashed line in Figure \ref{fig:10}).  Note
that the reorganization energy obtained from the fit is close to the
result of direct calculation from the second cumulant (eq \ref{eq:46},
Table \ref{tab:2}).

\begin{figure}
  \centering
   \includegraphics*[width=7cm]{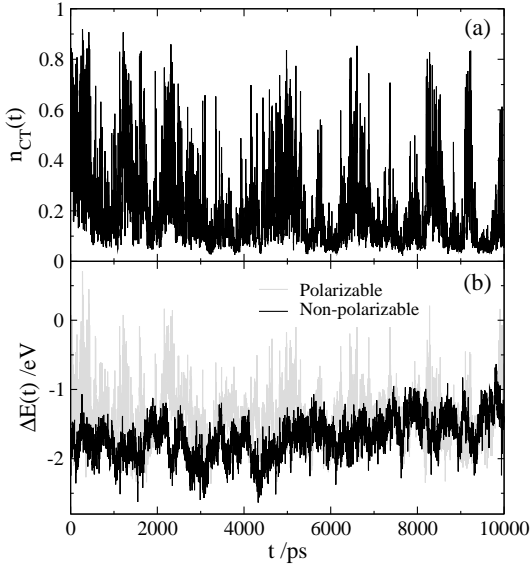}
   \caption{The trajectory of the population of the charge transfer
     state of P (a) and the trajectory of energy gap fluctuations (b)
     for non-polarizable (black line) and polarizable (gray line)
     special pair.  }
  \label{fig:11}
\end{figure}

The increase in the reorganization energy in the case of polarizable P
comes from fluctuations of the amount of charge transfer between the
covalent and ionized states of P (Figure \ref{fig:11}). It is clearly
seen from Figure \ref{fig:11} that energy gap fluctuations in excess
to those existing for non-polarizable P trace the fluctuations of
$n_{CT}$. Most of the excess reorganization energy comes from the
protein. The reorganization energy from water actually gets smaller
when polarizability is introduced, but the protein reorganization
energy is increased by a factor of four.

We do not currently have an established theoretical algorithm of how
to calculate the non-parabolic free energy surfaces of electron
transfer involving highly polarizable donor-acceptor states when
ergodicity is broken.  In the absence of a theoretical formalism, we
have turned to simulations. Figure \ref{fig:10} shows the free energy
surface produced from simulations by sliding the observation window of
the length 4 ps along the trajectory and then averaging all the
histograms produced from each window after shifting them to a common
probability maximum. The normalized distribution produced in this way
is then used to plot the non-ergodic free energy curve shown in Figure
\ref{fig:10}. In contrast to distributions obtained with the
non-polarizable simulation protocol, the non-ergodic surface turns out
to be non-parabolic. We do not presently have a good explanation of
this observation.

\subsection{Charge-transfer rates}
\label{sec:3-3}
\begin{figure}
  \centering
  \includegraphics*[width=9cm]{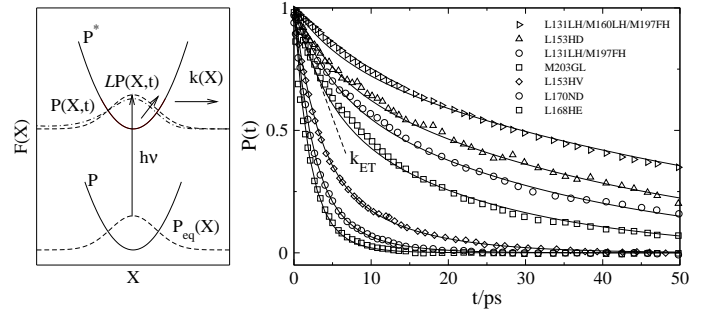}
  \caption{Left panel: Photoexcitation of the special pair lifting the
    equilibrium distribution (dashed line) from the ground state to
    the electronically excited state. This excitation starts the decay
    of the population through the Gaussian sink $k(X)$, along with
    the one-dimensional diffusion given by the Fokker-Planck operator
    $LP(x,t)$.  The dash-dotted line indicates depletion of the
    population at the side of the sink resulting in a slowing down of
    the population relaxation and in overall non-exponential
    kinetics. Right panel: Population decays of mutants of the
    reaction center of \textit{Rhodobacter sphaeroides} taken from ref
    \onlinecite{Wang:07} (points) and fits of $P(t)$ to the diffusion-reaction
    model (solid lines).  The legend in the right panel specifies mutants altering
    the local electrostatic potential at the special
    pair.\cite{Haffa:02} The dashed line marked $k_{\text{ET}}$ shows
    the initial population decay with the electron transfer rate
    constant according to eq \ref{eq:43}.}
  \label{fig:12}
\end{figure}

The decay of the population $P(t)$ of the photoexcited special pair is
known to be non-exponential.\cite{Vos:91,Du:92,Wang:07} Recent
explanations of this observation\cite{Wang:07,Chaudhury:07} have cast
the problem in terms of the Fokker-Planck kinetics with a Golden Rule
reaction sink, similar to formalisms developed in the past by Agmon
and Hopfield\cite{Agmon:83} and Sumi and Marcus.\cite{Sumi:86} This
theoretical algorithm offers the following physical picture. At the
initial time $t=0$, a laser flash lifts the equilibrium population
$P_{\text{eq}}(X)$ of the ground P to the excited state P$^*$ (dashed
line in the left panel in Figure \ref{fig:12}). At this moment, the
state P$^*$ is fully occupied, $P(0)=1$. The system can decay to the
charge-separated state with the frequency $\omega_e$ (eq \ref{eq:3}) at the
activated state $X=0$ thus depleting $P(t)$ and changing
$P_{\text{eq}}(X)$ to $P(X,t)$ (dash-dotted line in the left panel in
Figure \ref{fig:12}). At the initial time, $P(X,t)\simeq P_{\text{eq}}(X)$,
and the decay is determined by the equilibrium rate $k_{\text{ET}}$
given by eq \ref{eq:2}.  However, as the population of the activated
state $X=0$ depletes from that given by the Boltzmann distribution,
the continuation of the reaction requires a diffusional supply of the
population to the activated state. The result is a slower population
decay and effectively multiexponential kinetics.  Given that the
activation barrier is small for primary charge separation (Figure
\ref{fig:9}), the diffusional regime kicks in at the early stage of
the reaction leading to observable deviations from monoexponentiality.

Two complications need to be recognized in applying this
type of diffusion-reaction kinetics to the problem of primary charge
separation. The first complication, well-recognized in studies of the
dynamic solvent effect on electron transfer in small
molecules,\cite{Zhu:91,Gayathri:96} is related to the fact that the
Stokes shift dynamics are non-exponential, in particular in its initial
Gaussian stage. The common approach to the problem, going back to the
Sumi-Marcus formalism,\cite{Sumi:86} is to split the overall energy
gap $X$ into a fast, $x_f$, and slow, $x$, components, $X=x_f +
x$. The evolution operator along the reaction coordinate $X$ is then
averaged over the equilibrium distribution of the fast component,
resulting in a diffusion-reaction equation for the population along
the slow reaction coordinate $x$ only:
\begin{equation}
  \label{eq:21}
   \partial P(x,t)/\partial t = \left[ L(k_{ET},x) - k(x) \right] P(x,t)
\end{equation}
In this equation, $L(k_{ET},x)$ is a diffusional operator
\begin{equation}
  \label{eq:41}
    L(k_{ET},x) = D(k_{ET}) \frac{\partial}{\partial x} \left[\frac{\partial}{\partial x} 
        + \beta  \frac{\partial G(k_{ET},x)}{\partial x}\right]  
\end{equation}
describing the Fokker-Planck dynamics in the potential given by the
electron-transfer free energy surface. For multi-exponential decay, a
time-dependent diffusion constant can be used for the harmonic
potential $G(k_{\text{ET}},x)$,\cite{Okuyama:86} while an effective
relaxation time $\tau_{\text{eff}}$ needs to be defined for a general
potential. Following Hynes,\cite{Hynes:86} this relaxation time is
defined here in terms of a weighted sum of the corresponding rates of
exponential relaxation. For a bi-exponential long-time tail in eq
\ref{eq:15}, one gets
\begin{equation}
  \label{eq:42}
  \tau_{\text{eff}}^{-1} =  
     \left(\lambda_1^{\text{C}}\tau_1^{-1} + \lambda_2^{\text{C}} \tau_2^{-1}\right)\big/ \left(\lambda_1^{\text{C}}+\lambda_2^{\text{C}}\right)
\end{equation}
The diffusion constant in eq \ref{eq:41} then becomes
$D(k_{ET})=2k_{\text{B}}T\lambda^{\text{C}}(k_{ET})/\tau_{\text{eff}}$. Finally the rate constant
$k(x)$ in eq \ref{eq:21} is the Golden Rule rate averaged over the
equilibrium distribution of the fast relaxation component
\begin{equation}
  \label{eq:44}
  k(x) = \omega_e\sqrt{\lambda_s/ (\lambda_G^{\text{C}}+\lambda^{\text{ind}}) } \exp\left[ - \beta
    \frac{(\langle \Delta E \rangle  - x)^2
    }{4(\lambda_G^{\text{C}}+\lambda^{\text{ind}}) } \right]
\end{equation} 
where $\lambda_G^{\text{C}}$ is the fast Gaussian component of decay in eq
\ref{eq:15}, $\lambda^{\text{ind}}$ is the induction reorganization energy,
and $\langle\Delta E\rangle$ is given by eq \ref{eq:4}.

Most studies applying this formalism in the past have assumed that the
overall rate of diffusional reaction, i.e.\ the rate of arriving at
the transition state $X=0$ from the bottom of the potential well, is
much smaller than the relaxation rate of any nuclear mode coupled to
electron transfer. This is obviously not true in our case, and
non-ergodicity corrections, already introduced into eqs
\ref{eq:21}--\ref{eq:44}, are required. These corrections come in the
form of the free energy surface $G(k_{\text{ET}},x)$ depending on the
rate $k_{\text{ET}}$ (eq \ref{eq:35}), as well as the diffusion
coefficient $D(k_{\text{ET}})$ depending on the non-ergodic
reorganization energy $\lambda^{\text{C}}(k_{\text{ET}})$. Therefore, any solution of
the dynamic diffusion-reaction equation should produce a closure for
$k_{\text{ET}}$ and then solved by repeated
iterations.\cite{DMjpcb3:06} Equation \ref{eq:21} was solved employing the
generalized moment expansion of Nadler and Marcus\cite{Nadler:87} to
produce $k_{\text{ET}}$ as the initial population decay (right panel
in Figure \ref{fig:9})
\begin{equation}
  \label{eq:43}
        k_{\text{ET}}= - \frac{d\ln {P(t)}}{dt} \bigg|_{t\to0}
\end{equation}
where $P(t)=\int P(x,t) dx$. This condition establishes the closure for
the self-consistent calculation of $k_{\text{ET}}$ by repeated
solutions of eq \ref{eq:21}. The free energy surface is then obtained
by a horizontal shift of eq \ref{eq:53},
$G(k_{\text{ET}},x)=x^2/[4\lambda(k_{\text{ET}})]$.

The approach outlined here results in a good agreement with
experimental population decays for a number of mutants reported by
Wang \textit{et al}\cite{Wang:07} (Figure \ref{fig:12}) with the input
parameters produced by S1 simulation protocol.  Also notice that the
rate constant in the sink term in eqs \ref{eq:21} and \ref{eq:44} is
purely classical and does not incorporate quantum vibrations.  For
reactions in the inverted region, quantum Franck-Condon vibrational
overlaps provide additional vibronic channels for electronic
transitions.\cite{Walker:92} Primary charge separation appears to
operate in the normal region\cite{Haffa:02} (Figure \ref{fig:8}) and
quantum vibrations can be omitted. Notice, however, that classical
phonon modes have been included into the fast Gaussian and induction
components of the reorganization energy. 

\begin{figure}
  \centering
  \includegraphics*[width=7cm]{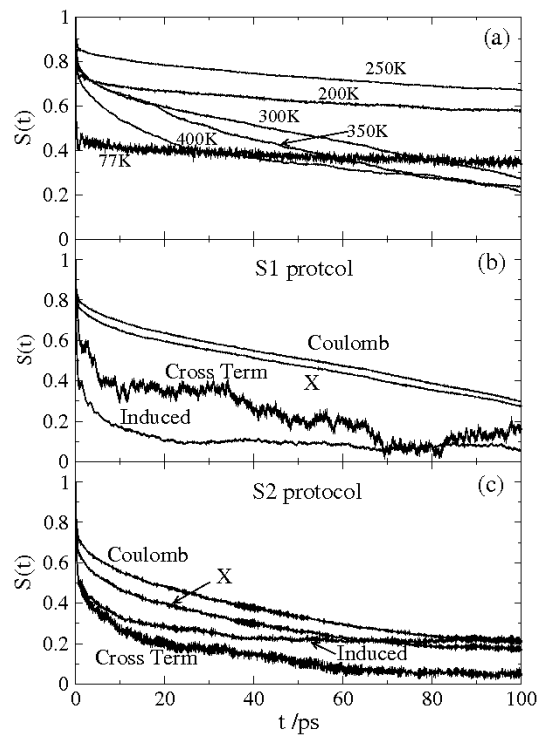}
  \caption{Stokes shift correlation function of primary charge
    separation obtained from 5-15 ns MD trajectories in S1 and S2
    simulation protocols. $S(t)$ at different temperatures with
    non-polarizable special pair are shown in (a).  In (b), the
    overall Stokes shift correlation function at 300 K (marked as $X$)
    is compared to its components from Coulomb and induction
    interactions, along with the Coulomb/induction cross term (eq
    \ref{eq:22}).  In (c), the same separation into components of
    $S(t)$ is shown for simulations with the polarizable special pair
    at 300 K. }
  \label{fig:13}
\end{figure}

The Stokes shift correlation function necessary to calculate the
non-ergodic reorganization energy from eq \ref{eq:18} at each
iteration step in eq \ref{eq:21} was taken from MD simulations of the
reaction complex (Figure \ref{fig:13}). Several important observations
follow from examining Figure \ref{fig:13}.  The ballistic component of
the decay, arising from ballistic motions of water and quasi-lattice
vibrations of the protein matrix, is significantly
diminished\cite{Abbyad:07} compared to the Stokes shift dynamics of
small chromophores in water.\cite{Jimenez:94} Indeed, the sum of the
Gaussian component of the Coulomb reorganization energy and the
induction reorganization energy, both responsible for the fast decay,
is below 20\% of the overall solvent reorganization energy $\lambda_s$. This
fact is critical for the analysis of non-ergodic free energy surfaces
of electron transfer as the fast relaxation component is essentially
the only part of nuclear reorganization which is not dynamically
arrested on the short time-scale of charge separation (see below).

The exponential tail of the Stokes shift decay becomes slower with
cooling, as expected. The effective relaxation time
$\tau_{\text{eff}}(T)$ can be calculated from the fitted exponential
relaxation times according to eqs \ref{eq:50} and \ref{eq:42} and fits
well by an Arrhenius function ($200\leq T\leq 400$ K) with the activation
energy of $E_{\tau}= 1212$ K. This activation barrier is close to the
value $\simeq 1060$ K reported for the long-time tail of the fluorescence
decay of an optical probe bound to a protein.\cite{Sahu:06} This
activation barrier was assigned to local segmental motions of the
protein coupled to the hydration layer. The long tail in the Stokes
shift correlation function observed here is, however, shorter than
that reported in ref \onlinecite{Sahu:06} and is in fact close to the
slow protein-water dynamics with the characteristic time of $\simeq 90$ ps
recently reported from Stokes shift data in ref \onlinecite{LiT:07}.

The combination of the Arrhenius temperature dependence with the low
activation energy points to the link between exponential Stokes shift
relaxation and $\beta$ relaxation of the protein/water system.  Previous
measurements of $\alpha$ relaxation in hydrated proteins have consistently
shown much larger effective activation barriers of the order
6000--9000 K,\cite{Fenimore:04,Markelz:07} in addition to the breaking
of the Arrhenius law in a broad temperature range. We therefore
conclude that primary charge separation is coupled to two nuclear
modes: Gaussian ballistic/phonon motions and exponential $\beta$
relaxation. The relaxation time of the former turns out be close to
0.1 ps\cite{Treutlein:92} and is essentially independent of
temperature.  We note in this regard that anharmonic protein
displacements shown in Figure \ref{fig:7} are also linked to $\beta$
relaxation.\cite{Fenimore:04} The decoupling of the Stokes shift
dynamics of the primary charge separation from $\alpha$ relaxation is
distinct from the situation commonly seen for solvation dynamics of
small solutes\cite{Reynolds:96} and, among other things, implies that
dielectric $\alpha$-relaxation data, routinely used to calculate solvation
dynamics of small chromophores,\cite{Bagchi:91} have little to do with
the dynamics of primary charge separation.

\begin{table*}
\caption{{\label{tab:3}}Solvent reorganization energies (eV) and their
  water and protein components from MD simulations. The reorganization
energies depending on the reaction rate were obtained from Stokes
shift dynamics according to eq \ref{eq:18}. $k_{\text{ET}}$ refers to
the charge separation rate at the corresponding temperature. }
\begin{tabular}{cccccccc}
\hline 
Protocol &  T/K &  $\lambda_s$ & $\lambda_{s}(k_{\text{ET}})$  &  
$\lambda_{\text{w}}$ &$\lambda_{\text{w}}(k_{\text{ET}})$ &  
$\lambda_{\text{prot}}$ & $\lambda_{\text{prot}}(k_{\text{ET}})$ \\
\hline
S1 &  77 &  0.421  &  0.266  &   0.191  &  0.072   &   0.245  &  0.210 \\
   & 200 &  0.903  &  0.257  &   0.756  &  0.106   &   0.251  &  0.201 \\
   & 250 &  1.955  &  0.261  &   1.639  &  0.126   &   0.419  &  0.235 \\
   & 300 &  1.598  &  0.454\footnotemark[1]  &   1.124  &  0.124   &   0.466  &  0.260 \\
   & 350 &  2.239  &  0.657  &   1.246  &  0.202   &   1.407  &  0.611 \\
   & 400 &  1.335  &  0.639  &   0.866  &  0.259   &   0.797  &  0.451 \\
S2 & 300 &  2.513  &  1.276  &   1.454  &  0.188   &   1.385  &  0.970 \\
\hline
\end{tabular}
\footnotetext[1]{Direct fits of the experimental population decays to the
  Sumi-Marcus model considering the average energy gap and the
  reorganization energy as fitting parameters gave 
  $\lambda_s=0.350$ eV and $\langle\Delta E\rangle =0.150$ eV.\cite{Wang:07} }  
\end{table*}

Stokes shift dynamics allow us to calculate the non-ergodic
reorganization energies entering the reaction rates and population
decays.  Table \ref{tab:3} reveals yet another important mechanistic
aspect. It shows that the reorganization energy of water is
significantly cut off by the dynamical arrest. Reorganization of fast,
anharmonic quasi-lattice vibrations of the protein emerges from the
water dominance in the thermodynamic limit, acting as the leading mode
driving electronic transitions on the picoseconds scale.

\begin{figure}
  \centering
  \includegraphics*[width=7cm]{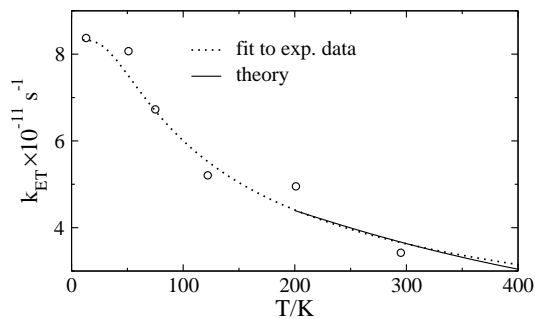}
  \caption{Temperature dependence of the rate of primary charge
    separation from experiments by Fleming \textit{et
      al}\cite{Fleming:88} (points) and from calculations of the rate
    using the gas-phase energy $\Delta E^{\text{gas}}=1.86$ eV and $V=41.5$
    cm$^{-1}$ adjusted to reproduce the rate at $T=300$ K (solid
    line).  The dotted line represents the fit of experimental data to
    an empirical equation suggested in ref \onlinecite{Fleming:88}.}
  \label{fig:14}
\end{figure}

We now turn back to the calculation of the rates of primary charge
separation. Two types of laboratory experiments are most relevant to
our discussion. The first are the measurements by Fleming and
co-workers\cite{Fleming:88} of charge separation rates in a broad
range of temperatures between helium 5 K and room temperature, 300 K.
The experimental observation, which has puzzled theorists ever since,
is a very gentle decay of the electron transfer rate over the whole
temperature range (open points in Figure \ref{fig:14}). This result is
apparently inconsistent with any conceivable temperature dependence of
equilibrium nuclear solvation energies, even if activationless
electronic transition is realized at some intermediate temperature.
The second set of experimental results, reported by Allen and Woodbury
and co-workers,\cite{Haffa:02,Wang:07} provides population decays of
P$^*$ in a carefully engineered set of mutants altering the
electrostatic potential at the location of the special pair. A
surprising result of these experiments was the realization that the
wild-type reaction center falls off the top of the energy gap law into
the normal region of electron transfer.\cite{Haffa:02}

Our current calculations, based on the input from MD simulations and
the concept of solvation non-ergodicity, are capable of reproducing
the experimental decay curves $P(t)$ for the whole set of mutants
studied by Allen and Woodbury (Figure \ref{fig:12}). In the fit, the
electron transfer matrix element was obtained from the 300 K rate of
the wild-type reaction center and the inhomogeneous component of the
average energy gap $\Delta E^{\text{C}}_{\text{inh}}$ was varied among the
mutants (electrostatic mutations\cite{Haffa:02}).  The fitted
variation of $\Delta E^{\text{C}}_{\text{inh}}$ is consistent with changes
in the midpoint electrochemical potential upon the mutation (Appendix
\ref{secC}). In order to further test the consistency of these results
with the experimental database, one needs to prove that the
experimental rates at different temperatures\cite{Fleming:88} can be
obtained with the set of parameters used to fit the mutagenesis data.
These results are shown in Figure \ref{fig:14} with the details of
calculations given in Appendix \ref{secC}.

Proper account of the temperature variation of the parameters entering
the activation barrier is important in reproducing the observed rates.
The main factor here is the temperature dependence of the induction
shift, which is well converged in our simulations and slopes
positively with increasing temperature (Table \ref{tab:1}).
Unfortunately, the accuracy of the current simulations does not allow
us to address the temperature dependence of the Coulomb components of
the energy shift and reorganization energy since their changes in the
interval of temperatures studied are within the uncertainties of
numerical simulations. Our previous experience with another
photosynthetic protein, plastocyanin, suggests that the length of the
simulated trajectories needs to be extended up to at least 20 ns for a
reliable estimate of the temperature slope,\cite{DMjcp2:08} which is
beyond our current computational capabilities. Therefore, in order to
assign realistic slopes to the Coulomb components of the free energy
barrier, we used our previous observation\cite{DMcp:06,DMjpca1:06}
that the results of many calculations and experiments on small solutes
in polar solvents give the logarithmic slope of the Coulomb
reorganization energy in the range $\Delta \lambda^C/ \lambda^C \simeq
-(2-3)\times10^{-3}\Delta T$. The Coulomb reorganization energy was
then assigned the slope of $\Delta \ln \lambda^C/ \Delta T
=-1.3\times10^{-3}$ K$^{-1}$ and, based on its relative magnitude, the
Coulomb component of the average energy gap was given the slope of
$\Delta (\ln \Delta E^C)/ \Delta T = 5.2 \times 10^{-4}$ K$^{-1}$ (see
Table \ref{tabC:1} in Appendix \ref{secC}). These assignments do not
affect our results much since a close fit can also be obtained by
assuming these these two parameters are temperature-independent. We finally note
that the problem of the temperature dependence of the reaction
parameters, in particular the driving force, is not free of
controversy. Opposite signs of reaction entropy have been obtained for
different charge-transfer reactions\cite{Volk:98,Edens:00} in the
reaction center and temperature-independent parameters are routinely
used in the analysis.\cite{GunnerDutton:89,Sumi:01}

\begin{figure}
  \centering
  \includegraphics*[width=7cm]{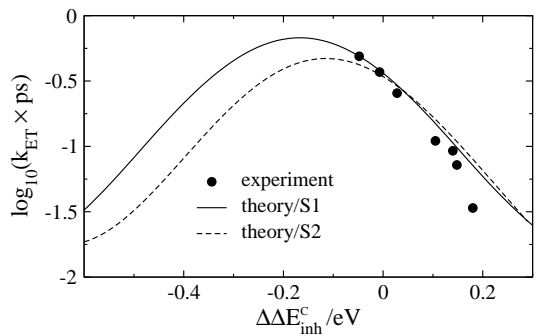}
  \caption{Charge separation rate vs the variation of the average
    donor-acceptor energy gap produced by mutagenetic substitution
    (points\cite{Wang:07}).  The lines are obtained by horizontal
    shifts of the non-ergodic parabolas of the initial charge
    separation state obtained in S1 protocol (solid line, Figure
    \ref{fig:9}) and in S2 protocol (dashed line, Figure
    \ref{fig:10}).  In experiment, mutagenetic substitution varies the
    inhomogeneous part of the Coulomb component of the average
    vertical gap and therefore that parameter marks the horizontal
    axis, $\Delta \Delta E^{\text{C}}_{\text{inh}}=0$ corresponds to the
    wild-type reaction center.  }
  \label{fig:15}
\end{figure}

The non-ergodic free energy surface of electron transfer (narrower
surface in Figure \ref{fig:10}) obtained from simulations with a
polarizable special pair can also be used to fit the experimental
reaction rate at 300 K. This fit results in the gas-phase gap of $\Delta
E^{\text{gas}}=1.57$ eV used in Figure \ref{fig:10} to plot the
free-energy surfaces. This value of the gas-phase gap yields the
average energy gap of $\langle\Delta E\rangle = 0.150$ eV, consistent with the
experimental evidence\cite{Haffa:02} and previous fits of the rates by
Wang \textit{et al}.\cite{Wang:07} The non-ergodic reorganization
energy obtained from fitting the curvature at the minimum of the
$G(k_{\text{ET}},X)$ curve turns out to be 0.39 eV, close to the value
of 0.46 eV reported for S1 simulations in Table \ref{tab:3} and the
value of 0.35 eV reported by Wang \textit{et al}.\cite{Wang:07} The
non-ergodic free energy curves from Figures \ref{fig:9} and
\ref{fig:10} are used to construct the energy gap law of electron
transfer plotted against the variation of the inhomogeneous
electrostatic potential of the protein, as was done in mutagenesis
experiments.\cite{Haffa:02} We find that polarizable and
non-polarizable simulations result in close shapes of the energy-gap
law.

\section{Discussion}
\label{sec:4}

\subsection{Mechanism of electron transfer activation}
The extensive MD simulations of the bacterial reaction center combined
with formal modeling have allowed us to look closely at the nuclear
modes driving electronic transitions and their energetic balance in
the reaction activation barrier. Several qualitative results have
emerged from our analysis. From the viewpoint of the relative
participation of different types of interaction potentials, we have
shown that induction and Coulomb forces give comparable contributions
to the average energy gap, while Coulomb interactions tend to dominate
the reorganization energy of electron transfer. A significant finding
of this study is the realization that, on the nanosecond time-scale
achievable by numerical simulations, the reorganization energies and
shifts are quite significant, much larger than had been anticipated so
far. The understanding that most of this nuclear solvation is
dynamically frozen on the time-scale of the reaction then became
critical for the quantitative description of the observable
rates. While water dominates the reorganization energy on the
nanosecond time-scale, most of this solvation freezes on the
picosecond reaction time-scale, and protein vibrations emerge as the
main nuclear mode driving electronic transition. Still, there is a
noticeable component of water reorganization, originating from the
ballistic Gaussian decay of the Stokes shift correlation function,
left even on the picosecond time-scale (Table \ref{tab:3}).

\begin{figure}
  \centering
  \includegraphics*[width=7cm]{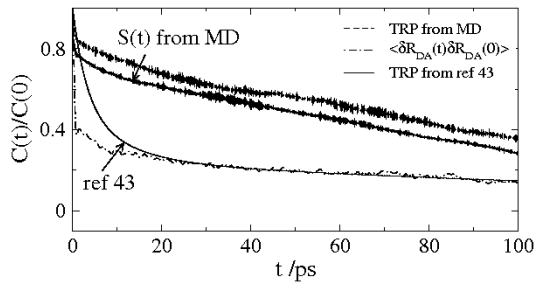}
  \caption{Total Stokes shift correlation function of primary charge
    separation compared to the Stokes shift correlation of reaction
    center tryptophans (dashed line) and to the normalized
    self-correlation function of the fluctuations in the
    donor-acceptor distance $R_{\text{DA}}(t)$ between the special
    pair and bacteriochlorophyll (dash-dotted line). The lower solid
    line shows the correlation function of tryptophan absorbance band
    taken from ref \onlinecite{Wang:07}. }
  \label{fig:16}
\end{figure}

Once the protein is identified as the major heat reservoir operating
on the picosecond time-scale of the reaction, one can try to identify
a particular mode most strongly coupled to the transferred electron
and driving the electronic transition. Several answers to this
question have been proposed in the past. Wang \textit{et
  al}.\cite{Wang:07} suggested to use transient changes in tryptophan
absorbance at 280 nm to monitor the electron transfer kinetics.  In
this approach, photoexcited tryptophan serves as a time-resolved probe
of the ultrafast nuclear rearrangement of the protein matrix with the
hope that the dynamics recordered by spectroscopy will match the
Stokes shift dynamics unreachable by spectroscopic techniques. Since
both types of information are available from our simulations, we have
tested this hypothesis here.

Figure \ref{fig:16} compares the Stokes shift dynamics of primary
charge separation with the Stokes shift dynamics of tryptophan
averaged over all tryptophans in the reaction center protein.
Although these two match each other reasonably well, the experimental
trace\cite{Wang:07} shown in the same plot is quite different having,
in particular, a much faster initial decay.\cite{com:TRP} It turns out
that this experimental trace matches quite well the autocorrelation
function of the donor-acceptor distance between the special pair and
the accessory bacteriochlorophyll cofactor B$_L$ (Figure \ref{fig:2},
also see Figures S1 and S2). The decay of this function is also caused
by protein vibrations suggesting that the experimental observation
traces one of the long-wavelength vibrational modes responsible for
large-scale protein motions, but not necessarily the modes
contributing primarily to the Stokes shift dynamics of primary charge
separation.

In fact, following an early proposal by Gehlen \textit{et
  al},\cite{Gehlen:94} Chaudhury \textit{et al.}\cite{Chaudhury:07}
recently suggested that donor-acceptor vibrations represent the mode
activating electronic transitions. Our current results do not support
this hypothesis. The dynamics of the donor-acceptor vibrations are
different from the Stokes shift dynamics. In addition, both the
self-correlation function of donor-acceptor distances and the
experimental trace of Wang \textit{et al}.\cite{Wang:07} produce too
large an amplitude of the initial decay which would make a larger
portion of nuclear solvation unfrozen on the time-scale of the
reaction, thus invalidating the analysis of the reaction rates (see
below). On the experimental side, related evidence shows the low
sensitivity of charge-recombination rates to high pressure (up to 345
MPa) which caused about 16\% of volume change of the
sample.\cite{Winsdor:89}

What has not been considered so far in the long history of modeling
primary charge separation is the possibility that a high
polarizability of the photoexcited special pair can significantly
modify the energetics of the reaction. Our simulations here are the
first attempt to understand the possible consequences of the gigantic
polarizability of the special pair for the charge-transfer energetics
and kinetics.  The polarizability of the special pair was modeled here
by the two-state model\cite{Lathrop:94,Cherepy:97} with the dynamic
adjustment of the population between charge-transfer and neutral
states of the primary pair along the simulation trajectory. The
two-state model has its obvious limitations and a possibility of a
broader spectrum of electronic states\cite{parsonWarshel:87,Chang:01}
can be considered in the future, along with improved modeling of the
temperature variation of the absorption spectrum of the special pair.
Nevertheless, the present simulations give the first insights into
what sort of changes to the energetics should be anticipated when the
polarizability has been taken into account.

What we have found here is consistent with previous studies of the
role of polarizability in the energetics of electron
transfer.\cite{DMjpca:99,DMjacs:03} The free energy surface of the
initial electron-transfer state involving a polarizable special pair
is significantly distorted compared to Marcus parabolas which we
obtained in the simulation protocol with the non-polarizable primary
pair (\textit{cf}.\ Figures \ref{fig:9} and \ref{fig:10}). The
reorganization energy, obtained as the variance of the energy-gap
fluctuations, is significantly enhanced compared to the case of
non-polarizable simulations, also in agreement with the previous
studies.\cite{DMjacs:03,DMjpca:04} The free energy curve could be
fitted with the analytical equations of the Q-model, which introduces a
non-parabolicity parameter in addition to the two-parameter
description of the Marcus model.

Although the free energy curve from the 15 ns trajectory shown in
Figure \ref{fig:10} is perhaps the most asymmetric electron transfer
free energy surface ever reported from numerical
simulations,\cite{DMjacs:03} most of this asymmetry is washed out by
the dynamical arrest of nuclear solvation on the reaction time-scale.
The free energy surface narrows down and approaches the Marcus
parabola on the 4 ps observation window (Figure \ref{fig:10}). In
fact, the rate of charge separation can be equally well described by
either polarizable or non-polarizable simulation data with a close
range of parameters and a close match between the resulting energy gap
laws (Figure \ref{fig:15}). It appears that what charge separation
probes on the picosecond observation window is a stripped surrogate of
the rich dynamics and thermodynamics of the protein/water
electrostatics on the time-scale of thermodynamic observables. Nature
has therefore played with dynamical time-scales to reduce these
complexities to a near-resonance electron transfer driven by ballistic
phonon motions.

\subsection{Rates of primary charge separation}
The ideas of non-ergodic nuclear solvation advocated here were tested
for consistency with experimental observations by calculating the
rates of charge separation as a function of
temperature,\cite{Fleming:88} and the kinetics of the population decay
depending on mutagenetic substitution.\cite{Wang:07} Since the
simulation protocol involving the non-polarizable special pair had
produced Marcus parabolas for the free energy surfaces, we were able
to use the diffusion-kinetic model advanced by Sumi and
Marcus\cite{Sumi:86} in order to calculate the population decay. The
main question we were asking here is whether the use of the same set
of fitting parameters (electron-transfer matrix element $V$ and the
gas-phase energy gap $\Delta E^{\text{gas}}$) would provide us with a
consistent description of both sets of experiments. We obtained a
positive result here (Figures \ref{fig:12} and \ref{fig:14}).

A fit of the population decay to the Sumi-Marcus diffusion-kinetic
model was also presented by Wang \textit{et al}.\cite{Wang:07} In
their analysis, the diffusion coefficient was taken to be
time-dependent in order to reflect the non-Markovian character of the
relaxation.\cite{Okuyama:86} Both the average energy gap and the
reorganization energy were considered as fitting parameters.  We found
that the use of the time-dependent diffusion coefficient is not a
necessity since the same data can be reproduced with an effective
diffusion coefficient extracted from Stokes shift correlation
functions. What distinguishes our analysis from their's is that the
solvent-induced shift of the average energy gap and the reorganization
energy are fixed by MD simulations, instead of used as fitting
parameters. The gas-phase gap and the electron transfer matrix element
were fitted to the rate at 300 K, but then, these parameters allowed
us to reproduce Fleming's data. With these restraints on the
parameters' magnitudes, there is very little room for adjusting the
two parameters. We also note that the results of the
calculations, and of our simulations of the charge-separated state
P$^+$-B$_L^-$, are consistent with the current state of experimental
evidence regarding the energetics of primary charge separation, as we
have discussed in sec \ref{sec:3-2}.

The negative slope of the charge separation rate with increasing
temperature has puzzled theorists for two decades, and has mostly been
approached by considering a temperature-dependent population of phonon
modes coupled to electron transfer.\cite{BixonJortner:99} Although our
simulations and conclusions are for the most part limited to high
temperatures greater than 200 K, explaining the negative temperature
slope of the rate in this region does not require vibronic coupling
models. We found the reaction rate to follow the temperature variation
of the induction component of the average energy gap which itself
becomes less negative with increasing temperature because of the
protein expansion.

We have confirmed the observation made by Haffa \textit{et
  al.}\cite{Haffa:02} that primary charge separation falls into the
normal region of electron transfer (Figures \ref{fig:9} and
\ref{fig:15}). This result was considered incompatible with the weak
temperature dependence of the rate, and a vibrational heating
mechanism\cite{Haffa:02,ParsonW:04,Parson:04} was suggested in order
to explain the positive $\langle \Delta E\rangle$. Our current calculations suggest
that Fleming's data can be reconciled with $\langle \Delta E\rangle \simeq 0.15$ eV for the
wild-type reaction center without assuming vibrational heating once
the temperature dependence of $\langle\Delta E\rangle$ is taken into account. The main
component of $\langle\Delta E\rangle$ responsible for its temperature dependence is the
shift by electronically instantaneous induction forces which do not
get dynamically frozen, but rather change due to a
temperature-affected alteration of protein's refractive index (eq
\ref{eq:26}). Our simulations also suggest that the wild-type reaction
center is driven even further from the optimum activationless
energetics when temperature increases above the room temperature and
that the optimum activationless configuration is reached at around 200
K (Figure S3). We refrain from speculations on evolutionary
implementations of this result.

\subsection{Broader insights}
Electron transfer connects cofactors in energetic redox chains in
biology. Three parameters are generally believed to have the main
impact on the kinetics of electron hops: the redox potential, the
probability of tunneling, and the reorganization energy. The first one
is relatively well understood, and in many cases, accessible to
measurements. The distance decay of electron tunneling has attracted
significant attention of the theoretical\cite{Beratan:92,Skourtis:05}
and experimental\cite{Gray:05} communities in recent decades.
Although the importance of specific pathways in the polypeptide
structure vs the generic tunneling decay specified by the height of
the potential barrier is still actively
discussed,\cite{Gray:05,Moser:06,Jasaitis:07,Beratan:08} there is a
general consensus about the magnitudes of matrix elements involved and
the distance decay of the tunneling
probability.\cite{Gray:05,Goldsmith:05}

The last component of the biological electron transfer picture, the
reorganization energy, is probably least understood.  Although the
reorganization energy is the hallmark of the classical Marcus theory
of electron transfer,\cite{MarcusSutin} not much is known about both
its value and the microscopic modes responsible for reorganization in
protein\cite{Gray:05} and DNA\cite{Lewis:001} electron transfer. For
proteins, the experimental evidence mostly comes from kinetic
measurements of ruthenium-modified proteins introduced into the field
by Gray and co-workers,\cite{Skov:98} and some recent reports from
computer simulations.\cite{Cascella:06,Blumberger:06,Sulpizi:07}
Notice that computer simulations reported in the past were mostly
limited to either very short
trajectories\cite{Treutlein:92,Parson:98,Cascella:06} or estimates of
the reorganization energy from the Stokes
shift,\cite{Blumberger:06,Sulpizi:07} which does not necessarily
provide the correct value of the reorganization energy defined through
the variance of the energy gap.\cite{DMjpcb1:08} The uncertainties of
reorganization energy values have led Dutton and co-workers to
suggest\cite{Moser:06} a generic value of 0.7 eV for electronic
transitions between cofactors not exposed to water with the provision
that smaller values might be required for photosynthetic electron
transfer. The fits of the photosynthetic rates have been attempted
many times and extremely low values of the reorganization energies (as
low as 0.1 eV\cite{Volk:98,Sumi:01}), completely unthinkable in
light of our present simulations, have been reported in the literature.

Our present work gives a different perspective to the problem of the
activation barrier of electron transfer in proteins. We claim that the
range of reorganization energies fundamentally attainable in protein
electron transfer is very broad given that the overall reorganization
energy attained in our present and previous\cite{DMjpcb1:08}
simulations is much higher than it was previously anticipated ($\simeq 1.6$
eV for S1 protocol and $\simeq 2.5$ eV for S2 protocol). The question of
assigning the reorganization energy thus turns not into its
``generic'' value, but into the question of finding the
protein/solvent reorganization energy reachable on a given time-scale
of the reaction, when a certain portion of nuclear degrees of freedom
is dynamically frozen.
 
We could not identify any specific solvent and/or protein modes that
drive electron transfer. Instead, the energetics of electronic
transitions appear to be driven by some generic set of ballistic
modes which would probably characterize any heterogeneous solvent made
of a rigid core (protein) surrounded by a molecular polar solvent
(water). It also seems true that achieving both the reaction rate of
primary charge separation and its low temperature dependence allows
some, although not large, flexibility in the driving force ($\simeq 0.3$ eV
between photosynthetic bacteria\cite{Volk:98}). Where the specific
design of the reaction center appears to matter is in providing a
sufficient tunneling rate between closely separated cofactors.  This
part of the design turns out to be very essential since the fast rate
allows the natural photosynthesis to dynamically freeze nuclear
solvation, and to achieve low values of the reorganization parameters
characterized by weak temperature dependence (ballistic motions and
local density fluctuations). It might therefore turn out that ``Darwin
at the molecular scale''\cite{Moser:06} operates not that much with
redox potentials but, to a greater extent, with relaxation
time-scales.

\textbf{Supporting Information Available:} Atomic charges of the
bacteriochlorophyll cofactors from DFT calculations and Figures
S1--S3. This material is available free of charge via the Internet at
\verb|http://pubs.acs.org|.

\begin{acknowledgments}
  This work was supported by the NSF (CHE-0616646). CPU time was
  provided by ASU's Center for High Performance Computing. We are
  grateful to Prof.\ M.\ Marchi for sharing with us the force field
  parameters of bacteriochlorophylls developed in his group. We also
  thank Prof.\ Mark Ratner for inspiring discussions of our results.
\end{acknowledgments}

\appendix

\section{Simulation protocol}
\label{secA}
Amber 8.0 \cite{amber8} was used for all MD simulations and
minimizations. The initial configuration of the reaction center
complex was taken from a crystal structure of the purple bacterium
\textit{Rhodobacter sphaeroides}.\cite{1pcr94} The force fields of
bacteriochlorophylls, pheophytins, ubiquinones, and iron center were
taken from Marchi and coworkers.  \cite{MarchiParms02} The protocol
for the creation of solvated micelle was taken from ref
\onlinecite{Ceccarelli1:03} with the slight variations described
below.

First, it should be mentioned that the reaction center was built
without the carotenoid cofactor, since it was deemed unnecessary for the
photosynthesis function.\cite{Cherepy:97} The system was initially
setup by protonating all lone valences and assuming standard pKa
values at pH$=7$.  Next, conjugate gradient minimization was applied
for $3N$ steps to remove bad contacts introduced by protonation ($N$
is the total number of protein atoms). A detergent micelle was then
created.  The micelle was made by placing on a circle of 8 LDAO molecules
in the first quadrant at the $z=0$ plane, with the heads pointing to
the exterior and the tails pointing to the origin.  Symmetry
transformations were applied about $x$ and $y$ axes to make a ring
of 32 LDAO out of the first quadrant LDAO molecules.  The ring was
then copied and translated along the $z$-axis to create four more
rings, each 7 \AA{} apart.  The protein was then rotated to align the
quasi-C$_2$ axis with the $z$-axis, and translated so that the origin
overlapped with the protein's center of mass.  The rings were placed
around the protein making a tight fit, which covered almost the entire
alpha helix region.

To help to form a micelle, the protein was allowed to relax by
conjugate gradient minimization for another $3N$ steps while the LDAO
were kept in place using a harmonic positional restraining force of
200 kcal/\AA$^{2}$.  Then, the force was removed and the system was
slowly heated in a vacuum at a rate of 20 K/ps until 200 K.  After
this short time, the LDAO shell melted and collapsed into a tight
micellar structure around the reaction center complex.  The total
energy of the system, the sum of the van der Waals and electrostatic
energy, decreased during heating by several thousand kcal/mol
indicating the creation of a more stable structure.  This step was
different from Ceccarelli and Marchi's approach\cite{Ceccarelli1:03}
which required heating to 400 K for several hundred ps in order to
form a compact, equilibrated micelle.  Once the micelle was formed,
the system charge was neutralized with 6 sodium ions, while another 30
NaCl pairs were added to keep the system at an approximate 0.15 M salt
concentration.  Then, a total of 10,506 waters (10,503 in the
charge-separated state) were added to form a truncated octahedron of
the simulation cell.

Following the addition of the solvent and counterions, each system was
run through an additional equilibration procedure.  First, water was
allowed to relax along a conjugate gradient minimization for $3N$
steps, while the micellar protein was held fixed with a weaker
restraining force of 10 kcal/ \AA$^{2}$.  Next, the full solvated
micellar system was allowed to relax for another $3N$ steps to remove
any remaining bad contacts.  Following this minimization, the solvated
system was heated again from 0 K to the desired temperature for 30 ps
(NVT ensemble).  After temperature equilibration, the volume was
allowed to expand in a 2 ns NPT run, which stabilized in less than
200 ps.  Once the density was equilibrated, NPT production runs
lasting 5--15 ns (1--5 ns at $T=77$ K) were used to calculate the
averages.

A single 2 fs timestep for all MD simulations was employed, and
SHAKE\cite{shake77} was used to constrain covalent bonds to hydrogen
atoms.  For constant temperature and pressure, the system was coupled
to a Berendsen thermostat and barostat, respectively.  The long-range
electrostatic interactions were handled using a smooth particle mesh
Ewald summation with a $10$ \AA\ limit in the direct space
sum.\cite{pme95}

\section{Atomic charges and charge transfer within the 
special pair}
\label{secB}
The partial charges of the electron transfer cofactors are not
provided by the Amber force field and need to be taken from quantum
calculations.  Due to the large size of the molecules, we modified the
bacteriochlorophyll (Bchl) cofactors by replacing their phytyl side
chains with methyl groups.  The quantum calculations of these modified
molecules were performed using GAMESS(US)\cite{gamess} (B3LYP
DFT/3-21G) and converted to partial charges by CHELPG protocol, also
implemented in GAMESS. The charge distribution of atoms of the phytyl
chain was assumed to be the same for the neutral and charged cofactors
and was calculated using the Antechamber module from Amber which
employs the empirical AM1-BCC method. The full sets of atomic charges
(with phytyl chains) given in Table S2 (supporting information) was
used in the MD simulations.  Similarly, the distribution of charge in
the final charge-transfer state was obtained from DFT partial charges
of a negatively charged Bchl$^-$ anion radical and a positively
charged cation radical Bchl$^+$ (Table S2). The positive charge was
distributed unequally between the two Bchls of the special pair, with
2/3 of the positive charge residing on the L subunit (P$_L$) and 1/3
of the positive charge residing on M subunit (P$_M$), as suggested by
ENDOR studies of \textit{Rhodobacter sphaeroides}.\cite{Johnson:02} The
set of $\Delta q_k$ charges ($k$ runs over the atoms of the cofactors)
obtained by subtracting the atomic charges in the initial neutral
state from the ionized state were used to calculate the Coulomb part
of the donor-acceptor energy gap. The partial charges on the protein
atoms were taken from the Amber FF03 force field,\cite{amberFF03} and
the TIP3P force field\cite{tip3p:83} was used for the partial charges
of water.

We used Stark spectroscopy data by Lockhart and
Boxer\cite{Lockhart:88} as the starting point for determining the
parameters of the charge-transfer state of the photoexcited special
pair. The change in the absorption dipole moment within the special
pair is about $f_c \Delta \mu=$ 7 D larger than in an isolated
bacteriochlorophyll, where $f_c\simeq 1.2$ is the cavity field correction
factor. If this change of the dipole moment difference, measured at 77
K, is connected to the mixing between the covalent (P$_M$-P$_L)^*$ and
charge-separated, (P$^+_M$-P$^-_L)^*$ states of the exited special pair,
then this change in the dipole moment can be written as
\begin{equation}
\label{eq:B1}
  \Delta \mu = n_{\text{CT}} \Delta \mu_{\text{CT}}
\end{equation}
where $\Delta \mu_{\text{CT}}$ is the dipole moment of the fully ionized
state (P$_M^+$-P$_L^-)^*$ and $n_{\text{CT}}$ is the population of this
state at the given energy gap between the neutral and ionized
states. In terms of the two-state Hamiltonian, this population is
given as
\begin{equation}
  \label{eq:B2}
  n_{\text{CT}} = \frac{1}{2} - \frac{\Delta \epsilon }{2\Delta \omega_P}
\end{equation}
where $\Delta \epsilon$ is the difference between diabatic energies of the neutral
and ionized states (eq \ref{eq:32}) and $\Delta \omega_P$ is the adiabatic energy
gap between the eigenvalues of the two-state Hamiltonian
\begin{equation}
  \label{eq:B3}
  \Delta \omega_P =\left(\Delta \epsilon^2 + 4 J^2\right)^{1/2}
\end{equation}
Here, $J$ is the electronic coupling element between the neutral and
ionized states of the excited special pair and, following Lathrop and
Friesner,\cite{Lathrop:94} we assume that the charge-transfer state
(P$^+_M$-P$^-_L)^*$ is predominantly mixed with the lower excitonic state
of the dimer.

When two bacteriochlorophyll radicals, P$_L^-$ and P$_M^+$ are placed
at their crystallographic positions, the resulting dipole moment of
the fully ionized state is $\Delta \mu_{\text{CT}}=40.2$ D. This implies that
average charge mixing between the two states at 77 K is
$n_{\text{CT}}(77 \mathrm{K})=0.143$. In order to determine the
coupling parameter $J$ from this number we used the model vibronic
Hamiltonian of Friesner and co-workers which was shown to reproduce a
number of experimental properties (absorption, circular dichroism,
polarized absorption).\cite{Lathrop:94} In this model, the difference
of energies between the ionized charge-transfer and neutral states of
the special pair is 2800 cm$^{-1}$, which, combined with the
population of charge-transfer state, gives $J=979$ cm$^{-1}$ and $\Delta
\epsilon=1998$ cm$^{-1}$. The former value falls in between 600 cm$^{-1}$
used by Lathrop and Friesner\cite{Lathrop:94} and 1450 cm$^{-1}$ used
by Renger.\cite{Renger:04}

The electronic mixing between the neutral and ionized states of the
special pair will make its excited state more polarizable than the
ground state. The change in the polarizability associated with charge
transfer can be readily calculated from the two-state polarizability
model which gives
\begin{equation}
  \label{eq:B4}
  \Delta \alpha = 2\Delta\mu_{\text{CT}}^2 J^2/(\Delta \epsilon)^3
\end{equation}
With the parameters calculated above, this equation gives $\Delta \alpha=707$
\AA$^3$, consistent with $\Delta \alpha =460-745$ \AA$^{3}$ reported from fitting
the Stark spectra.\cite{Middendorf:93}

The energy gap $\Delta \epsilon$ was obtained in by fitting the spectra at 77
K\cite{Lathrop:94} and is not directly suitable for our simulations at
higher temperatures. The average energy gap between neutral and
ionized states is made by the gas-phase gap $\Delta \epsilon^{\text{gas}}$ and a
shift by polar and induction interactions with the protein/water
solvent (eq \ref{eq:32}). In order to extract this shift, we have run
a short (1 ns) MD simulation of the reaction center at 77 K from which
the solvent shift was determined to be $-0.974$ eV.  This number
allowed us to determine the gas-phase gap of $\Delta \epsilon^{\text{gas}} =
1.222$ eV which was used in the simulations of the polarizable special
pair. The simulations required modification of Sander module of AMBER
such that the instantaneous energy gap and special pair charges are
recalculated at each fifth time step of the MD run.

\section{Fitting experimental kinetic data}
\label{secC}
Our model was applied to two sets of experimental kinetic data, the
temperature dependence of the primary rate from Fleming \textit{et
  al}\cite{Fleming:88} and time-resolved decays of the population of
the photoexcited special pair from Wang \textit{et al}.\cite{Wang:07}
For the latter set of data, recordered at $T=300$ K, we used the
solvent reorganization energy from our MD simulations with the
non-ergodic correction extracted from the Stokes shift dynamics (eq
\ref{eq:18}).  Population decays were calculated by solving the
diffusion-reaction Fokker-Planck equations (eqs
\ref{eq:21}--\ref{eq:43}) for the mutants used in the experiment
(Table S1). In contrast to Wang \textit{et al}\cite{Wang:07} who used
three fitting parameters in their analysis, the reorganization energy
is fixed here by MD simulations and only the gas-phase energy gap $\Delta
E^{\text{gas}}$ and the electron transfer matrix element $V$ were
varied to fit the rate constant at 300 K for mutant L170ND, which is
very close to the wild type reaction center with the difference
between their mid-point potentials of only $-0.007$ eV.\cite{Wang:07}
This fit has resulted in $\Delta E^{\text{gas}} = 1.86$ eV and $V=41.5$
cm$^{-1}$ ($>60$ cm$^{-1}$ was identified for this parameter in ref 
\onlinecite{BixonJortner:89}).

\begin{figure}
  \centering
  \includegraphics*[width=7cm]{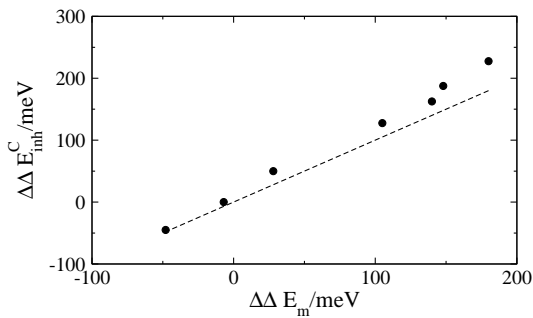}
  \caption{Correlation between the change of the inhomogeneous Coulomb
    energy gap $\Delta \Delta E^{\text{C}}_{\text{inh}}$ between
    mutants and wild-type reaction center and the corresponding
    changes in the midpoint redox potential $\Delta \Delta E_m$
    reported experimentally.\cite{Wang:07} The values of $\Delta
    \Delta E^{\text{C}}_{\text{inh}}$ are obtained from fitting the
    theoretical curves for the population decay to
    experiment\cite{Wang:07} (Figure \ref{fig:12}) while keeping the
    electron transfer matrix element and the gas-phase gap constant.
    The data shown in the plot are also listed in Table S1 (supporting
    information). The dashed line indicates the unitary slope to guide
    the eye. }
  \label{figC:1}
\end{figure}

Special pair mutants introduce electrostatic perturbations at the
location of two sandwiched bacteriochlorophylls without significantly
affecting the solvation component of the reaction Gibbs energy. This
implies the variation of the inhomogeneous Coulomb component of the
energy gap, $\Delta E^{\text{C}}_{\text{inh}}$. This component was varied in
fitting the experimental $P(t)$ curves of other mutants while keeping
the gas-phase gap and the electron transfer matrix element
constant. The variation of $\Delta E^{\text{C}}_{\text{inh}}$ with mutation
relative to the wild-type reaction center then closely follows changes
in the midpoint potential (Figure \ref{figC:1}).

\begin{table*}
\caption{{\label{tabC:1}} Parameters used to fit the charge-separation
  rate at 300 K and to produce the temperature dependence of the
  rate.$^a$  All parameters refer to the wild-type reaction center; 
  temperature derivatives taken at 300 K are in K$^{-1}$. }
\begin{tabular}{cccccc}
\hline
 $\Delta E^{\text{gas}}$, eV & $V$, cm$^{-1}$ &  $d\ln\Delta E^{\text{C}}/dT$  & 
                          $d\ln\Delta E^{\text{ind}}/dT$ &
                          $d\ln\lambda^{\text{C}}/dT$  & $d\ln\lambda^{\text{ind}}/dT$  \\
\hline
1.86 & 41.5 & $5.2\times10^{-4}$  &  $4.8\times 10^{-4}$ & $-1.3\times 10^{-3}$  &
              $1.1 \times 10^{-2}$ \\
\hline
\end{tabular}
\footnotetext[1]{Temperature dependence of the Stokes shift
  correlation function was produced by using the following parameters
  in eq \ref{eq:50}: $A_G$ =0.172, $\tau_G=0.1$, $A_1 = 0.063$, $\tau_1= 2.5$ ps,
  $\tau_2^0 = 2.55$ ps,  
  $E_{\tau} = 1212$ K with $\tau_2(T)=\tau_2^0 \exp[\beta E_{\tau}]$.  }
\end{table*}

The gas-phase shift $\Delta E^{\text{gas}}$ and the matrix element $V$,
obtained from fitting the rate at 300 K, were then used to calculate
the temperature dependence of the rate. This calculation is
complicated by the fact that all solvation energies and solvation
relaxation times depend on temperature. The MD simulations do not
provide sufficient accuracy to reliably estimate the temperature
change of the Coulomb part of the reorganization parameters. Their
temperature dependence was estimated from linear interpolations with
the slopes listed in Table \ref{tabC:1} (see also the discussion in
sec \ref{sec:3-3}).  The temperature dependence of the induction
component of the average gap is the main ingredient in reproducing the
slope of the rate correctly. This component is converged exceptionally
well in MD simulations which were used to produce $\Delta
E^{\text{ind}}(T)$.
 
The temperature dependence of the Stokes shift relaxation time makes
the non-ergodicity correction factor $f^{\lambda}(k_{\text{ET}})$
temperature-dependent as well. It turned out that only the relaxation
time $\tau_2(T)$ is strongly temperature-dependent in the Stokes shift
correlation function approximated by three components according to eq
\ref{eq:15}. The first two components were given constant values,
$\tau_G=0.1$ ps and $\tau_1=2.5$ ps, and the longest relaxation time was
given the Arrhenius temperature law $\ln \tau_2(T) = 0.936 + E_{\tau}/T $
with $E_{\tau}= 1212.3$ K.  The results of these calculations, shown in
Figure \ref{fig:14}, are in a good agreement with the data by
Fleming \textit{et al}.\cite{Fleming:88} in the range of temperatures
available to our simulations.

\bibliographystyle{achemso}

\bibliography{/home/dmitry/p/bib/chem_abbr,/home/dmitry/p/bib/photosynthNew,/home/dmitry/p/bib/glass,/home/dmitry/p/bib/et,/home/dmitry/p/bib/dm,/home/dmitry/p/bib/protein,/home/dmitry/p/bib/solvation,/home/dmitry/p/bib/bioet,/home/dmitry/p/bib/etnonlin,/home/dmitry/p/bib/dynamics,/home/dmitry/p/bib/simulations,/home/dmitry/p/bib/liquids,/home/dmitry/p/bib/pt,/home/dmitry/p/bib/glasset}

\providecommand{\refin}[1]{\\ \textbf{Referenced in:} #1}
\begin{thebibliography}{100}

\bibitem{Fleming:88}
Fleming,~G.~R.;\ \ Martin,~J.~L.;\ \ Breton,~J. \textit{Nature} \textbf{1988,}
  \textsl{333,} 190.

\bibitem{Hoff:97}
Hoff,~A.~J.;\ \ Deisenhofer,~J. \textit{Physics Reports} \textbf{1997,}
  \textsl{287,} 1.

\bibitem{Page:99}
Page,~C.~C.;\ \ Moser,~C.~C.;\ \ Chen,~X.~X.;\ \ Dutton,~P.~L. \textit{Nature}
  \textbf{1999,} \textsl{402,} 47.

\bibitem{Warshel:01}
Warshel,~A.;\ \ Parson,~W.~W. \textit{Quat. Rev. Biophys.} \textbf{2001,}
  \textsl{34,} 563.

\bibitem{Blankenship:02}
Blankenship,~R.~E. \textit{Molecular Mechanisms of Photosynthesis;} Blackwell
  Science: Williston, VT, 2003.

\bibitem{Page:03}
Page,~C.~C.;\ \ Moser,~C.~C.;\ \ Dutton,~P.~L. \textit{Curr. Opinion in
  Biology} \textbf{2003,} \textsl{7,} 551.

\bibitem{Zhang:98}
Zhang,~L.~Y.;\ \ Friesner,~R.~A. \textit{Proc. Natl. Acad. Sci. U.S.A.}
  \textbf{1998,} \textsl{95,} 13603.

\bibitem{Ivashin:98}
Ivashin,~N.;\ \ Kallebring,~B.;\ \ Larsson,~S.;\ \ Hansson,~O. \textit{J.\
  Phys.\ Chem.\ B} \textbf{1998,} \textsl{102,} 5017.

\bibitem{Nishioka:05}
Nishioka,~H.;\ \ Kimura,~A.;\ \ Yamato,~T.;\ \ Kawatsu,~T.;\ \ Kakitani,~T.
  \textit{J.\ Phys.\ Chem.\ B} \textbf{2005,} \textsl{109,} 1978.

\bibitem{Bixon:95}
Bixon,~M.;\ \ Jortner,~J.;\ \ Mechel-Beyerle,~M.~E. \textit{Chem. Phys.}
  \textbf{1995,} \textsl{197,} 389.

\bibitem{BixonJortner:99}
Bixon,~M.;\ \ Jortner,~J. \textit{Adv. Chem. Phys.} \textbf{1999,}
  \textsl{106,} 35.

\bibitem{Volk:98}
Volk,~M.;\ \ Aumeier,~G.;\ \ Langenbacher,~T.;\ \ Feick,~R.;\ \ Ogrodnik,~A.;\
  \ Michel-Beyerle,~M.-E. \textit{J. Phys. Chem. B} \textbf{1998,}
  \textsl{102,} 735.

\bibitem{Winter:03}
de~Winter,~A.;\ \ Boxer,~S.~G. \textit{J. Phys. Chem. A} \textbf{2003,}
  \textsl{107,} 3341-3350.

\bibitem{Noy:06}
Noy,~D.;\ \ Moser,~C.~C.;\ \ Dutton,~P.~L. \textit{Biochim. Biophys. Acta}
  \textbf{2006,} \textsl{1757,} 90.

\bibitem{Rossky:94}
Rossky,~P.~J.;\ \ Simon,~J.~D. \textit{Science} \textbf{1994,} \textsl{370,}
  263.

\bibitem{Jimenez:94}
Jimenez,~R.;\ \ Fleming,~G.~R.;\ \ Kumar,~P.~V.;\ \ Maroncelli,~M.
  \textit{Nature} \textbf{1994,} \textsl{369,} 471.

\bibitem{Reynolds:96}
Reynolds,~L.;\ \ Gardecki,~J.~A.;\ \ Frankland,~S. J.~V.;\ \ Maroncelli,~M.
  \textit{J. Phys. Chem.} \textbf{1996,} \textsl{100,} 10337.

\bibitem{Maroncelli:88}
Maroncelli,~M.;\ \ Fleming,~G.~R. \textit{J.\ Chem.\ Phys.} \textbf{1988,}
  \textsl{89,} 5044.

\bibitem{FlemingWolynes:90}
Fleming,~G.~R.;\ \ Wolynes,~P.~G. \textit{Physics Today} \textbf{1990,}
  \textsl{43,} 36.

\bibitem{Fenimore:04}
Fenimore,~P.~W.;\ \ Frauenfelder,~H.;\ \ McMahon,~B.~H.;\ \ Young,~R.~D.
  \textit{Proc. Natl. Acad. Sci.} \textbf{2004,} \textsl{101,} 14408.

\bibitem{Ediger:96}
Ediger,~M.~D.;\ \ Angell,~C.~A.;\ \ Nagel,~S.~R. \textit{J. Phys. Chem.}
  \textbf{1996,} \textsl{100,} 13200.

\bibitem{Bagchi:91}
Bagchi,~B.;\ \ Chandra,~A. \textit{Adv. Chem. Phys.} \textbf{1991,}
  \textsl{80,} 1.

\bibitem{Ranko:00}
Richert,~R. \textit{J. Chem. Phys.} \textbf{2000,} \textsl{113,} 8404.

\bibitem{Ngai:00}
Ngai,~K.~L. \textit{J. Non-Cryst. Sol.} \textbf{2000,} \textsl{275,} 7.

\bibitem{Yang:07}
Yang,~Y.;\ \ Lai,~W.-C.;\ \ Hsu,~S.~L. \textit{J. Chem. Phys.} \textbf{2007,}
  \textsl{127,} 054901.

\bibitem{Ngai:07}
Ngai,~K.~L.;\ \ Capaccioli,~S. \textit{J. Phys.: Condens. Matter}
  \textbf{2007,} \textsl{19,} 205114.

\bibitem{Marchi:93}
Marchi,~M.;\ \ Gehlen,~J.~N.;\ \ Chandler,~D.;\ \ Newton,~M. \textit{J. Am.
  Chem. Soc.} \textbf{1993,} \textsl{115,} 4178.

\bibitem{Schmidt:94}
Schmidt,~S.;\ \ Arlt,~T.;\ \ Hamm,~P.;\ \ Huber,~H.;\ \ N{\"a}gele,~T.;\ \
  Wachtveitl,~J.;\ \ Meyer,~M.;\ \ Scheer,~H.;\ \ Zinth,~W. \textit{Chem. Phys.
  Lett.} \textbf{1994,} \textsl{223,} 116.

\bibitem{Ogrodnik:94}
Ogrodnik,~A.;\ \ Keupp,~W.;\ \ Volk,~M.;\ \ Aumeier,~G.;\ \
  Michel-Beyerle,~M.~E. \textit{J. Phys. Chem.} \textbf{1994,} \textsl{98,}
  3432.

\bibitem{Holzwarth:96}
Holzwarth,~A.;\ \ Muller,~M. \textit{Biochemistry} \textbf{1996,} \textsl{35,}
  11820.

\bibitem{Yakovlev:00}
Yakovlev,~A.~G.;\ \ Shkuropatov,~A.~C.;\ \ Shuvalov,~V.~A. \textit{FEBS Lett.}
  \textbf{2000,} \textsl{466,} 209-212.

\bibitem{Zinth:05}
Zinth,~W.;\ \ Wachtveitl,~J. \textit{ChemPhysChem} \textbf{2005,} \textsl{6,}
  871.

\bibitem{Roberts:01}
Roberts,~J.;\ \ Holten,~D.;\ \ Kirmaier,~C. \textit{J. Phys. Chem. B}
  \textbf{2001,} \textsl{105,} 5575.

\bibitem{Warshel:80}
Warshel,~A. \textit{Proc. Natl. Acad. Sci.} \textbf{1980,} \textsl{77,}
  3105.

\bibitem{Lathrop:94}
Lathrop,~E. J.~P.;\ \ Friesner,~R.~A. \textit{J. Phys. Chem.} \textbf{1994,}
  \textsl{98,} 3056.

\bibitem{Reimers:03}
Reimers,~J.~R.;\ \ Shapley,~W.~A.;\ \ Hush,~N.~S. \textit{J.\ Chem.\ Phys.}
  \textbf{2003,} \textsl{119,} 3240.

\bibitem{Amerongen:00}
van Amerongen,~H.;\ \ Valkunas,~L.;\ \ van Grondelle,~R. \textit{Photosynthetic
  Excitons;} World Scientific: Singapore, 2000.

\bibitem{Middendorf:93}
Middendorf,~T.~R.;\ \ Mazzola,~L.~T.;\ \ Lao,~K.~Q.;\ \ Steffen,~M.~A.;\ \
  Boxer,~S.~G. \textit{Biochimica et Biophysica Acta} \textbf{1993,}
  \textsl{1143,} 223.

\bibitem{DMjpca:99}
Matyushov,~D.~V.;\ \ Voth,~G.~A. \textit{J. Phys. Chem. A} \textbf{1999,}
  \textsl{103,} 10981.

\bibitem{DMjacs:03}
Small,~D.~W.;\ \ Matyushov,~D.~V.;\ \ Voth,~G.~A. \textit{J. Am. Chem. Soc.}
  \textbf{2003,} \textsl{125,} 7470.

\bibitem{DMjpcb2:06}
Matyushov,~D.~V. \textit{J. Phys. Chem. B} \textbf{2006,} \textsl{110,} 10095.

\bibitem{1pcr94}
Ermler,~U.;\ \ Fritzsch,~G.;\ \ Buchanan,~S.~K.;\ \ Michel,~H.
  \textit{Structure} \textbf{1994,} \textsl{2,} 925.

\bibitem{Wang:07}
Wang,~H.;\ \ Lin,~S.;\ \ Allen,~J.~P.;\ \ Williams,~J.~C.;\ \ Blankert,~S.;\ \
  Laser,~C.;\ \ Woodbury,~N.~W. \textit{Science} \textbf{2007,} \textsl{316,}
  747.

\bibitem{Sumi:86}
Sumi,~H.;\ \ Marcus,~R.~A. \textit{J. Chem. Phys.} \textbf{1986,} \textsl{84,}
  4894.

\bibitem{Nadler:87}
Nadler,~W.;\ \ Marcus,~R.~A. \textit{J.\ Chem.\ Phys.} \textbf{1987,}
  \textsl{86,} 3906.

\bibitem{Zhu:91}
Zhu,~J.~J.;\ \ Rasaiah,~J.~C. \textit{J. Chem. Phys.} \textbf{1991,}
  \textsl{95,} 3325.

\bibitem{amber8}
Case,~D.~A.;\ \ Cheatham,~T.~E.;\ \ Darden,~T.;\ \ Gohlke,~H.;\ \ Luo,~R.;\ \
  Jr.,~K. M.~M.;\ \ Onufriev,~A.;\ \ Simmerling,~C.;\ \ Wang,~B.;\ \
  Woods,~R.~J. \textit{J.\ Comp.\ Chem.} \textbf{2005,} \textsl{26,} 1668.

\bibitem{Ceccarelli1:03}
Ceccarelli,~M.;\ \ Marchi,~M. \textit{J. Phys. Chem. B} \textbf{2003,}
  \textsl{107,} 1423.

\bibitem{Kirmaier:03}
Kirmaier,~G.;\ \ Laible,~P.~D.;\ \ Hindin,~E.;\ \ Hanson,~D.~K.;\ \ Holten,~D.
  \textit{Chem. Phys.} \textbf{2003,} \textsl{294,} 305.

\bibitem{DMacc:07}
Matyushov,~D.~V. \textit{Acc. Chem. Res.} \textbf{2007,} \textsl{40,} 294.

\bibitem{Kuharski:88}
Kuharski,~R.~A.;\ \ Bader,~J.~S.;\ \ Chandler,~D.;\ \ Sprik,~M.;\ \
  Klein,~M.~L.;\ \ Impey,~R.~W. \textit{J. Chem. Phys.} \textbf{1988,}
  \textsl{89,} 3248.

\bibitem{MarcusSutin}
Marcus,~R.~A.;\ \ Sutin,~N. \textit{Biochim. Biophys. Acta} \textbf{1985,}
  \textsl{811,} 265.

\bibitem{Pekar:46}
Pekar,~S.~I. \textit{JETPh} \textbf{1946,} \textsl{16,} 341.

\bibitem{Pekar:63}
Pekar,~S.~I. \textit{Research in electron theory of crystals;} USAEC:
  Washington, D.C., 1963.

\bibitem{Frohlich:54}
Fr{\"o}lich,~H. \textit{Adv. Phys.} \textbf{1954,} \textsl{3,} 325.

\bibitem{Feynman:55}
Feynman,~R.~P. \textit{Phys. Rev.} \textbf{1955,} \textsl{97,} 660.

\bibitem{DMjcp:95}
Matyushov,~D.~V.;\ \ Schmid,~R. \textit{J. Chem. Phys.} \textbf{1995,}
  \textsl{103,} 2034.

\bibitem{DMjcp2:05}
Matyushov,~D.~V. \textit{J. Chem. Phys.} \textbf{2005,} \textsl{122,} 084507.

\bibitem{Marcus:56}
Marcus,~R.~A. \textit{J. Chem. Phys.} \textbf{1956,} \textsl{24,} 966.

\bibitem{Raineri:99}
Raineri,~F.~O.;\ \ Friedman,~H.~L. \textit{Adv. Chem. Phys.} \textbf{1999,}
  \textsl{107,} 81.

\bibitem{Kornyshev:96}
Kornyshev,~A.~A.;\ \ Sutmann,~G. \textit{J. Chem. Phys.} \textbf{1996,}
  \textsl{104,} 1524.

\bibitem{DMcp:06}
Milischuk,~A.~A.;\ \ Matyushov,~D.~V.;\ \ Newton,~M.~D. \textit{Chem. Phys.}
  \textbf{2006,} \textsl{324,} 172.

\bibitem{Landau5}
Landau,~L.~D.;\ \ Lifshits,~E.~M. \textit{Statistical Physics;} Pergamon Press:
  New York, 1980.

\bibitem{Ovchinnikov:69}
Ovchinnikov,~A.~A.;\ \ Ovchinnikova,~M.~Y. \textit{JETPh} \textbf{1969,}
  \textsl{29,} 688.

\bibitem{Parson:98}
Parson,~W.~W.;\ \ Chu,~Z.~T.;\ \ Warshel,~A. \textit{Biophys. J.}
  \textbf{1998,} \textsl{74,} 182.

\bibitem{Palmer:82}
Palmer,~R.~G. \textit{Adv. Phys.} \textbf{1982,} \textsl{31,} 669.

\bibitem{DMjcp2:06}
Ghorai,~P.~K.;\ \ Matyushov,~D.~V. \textit{J. Chem. Phys.} \textbf{2006,}
  \textsl{124,} 144510.

\bibitem{Lockhart:88}
Lockhart,~D.~J.;\ \ Boxer,~S.~G. \textit{Proc. Natl. Acad. Sci. USA}
  \textbf{1988,} \textsl{85,} 107.

\bibitem{Haran:96}
Haran,~G.;\ \ Wynne,~K.;\ \ Moser,~C.~C.;\ \ Dutton,~P.~L.;\ \
  Hochstrasser,~R.~M. \textit{J. Phys. Chem.} \textbf{1996,} \textsl{100,}
  5562.

\bibitem{Wynne:96}
Wynne,~K.;\ \ Haran,~G.;\ \ Reid,~G.~D.;\ \ Moser,~C.~C.;\ \ Dutton,~P.~L.;\ \
  Hochstrasser,~R.~M. \textit{J. Phys. Chem.} \textbf{1996,} \textsl{100,}
  5140.

\bibitem{Arnett:99}
Arnett,~D.~C.;\ \ Moser,~C.~C.;\ \ Dutton,~P.~L.;\ \ Scherer,~N.~F. \textit{J.
  Phys. Chem. B} \textbf{1999,} \textsl{103,} 2014.

\bibitem{Larsson:90}
Larsson,~S.;\ \ K{\"a}llbring,~B. \textit{Int. J. Quant. Chem.: Quant. Biochem.
  Symp.} \textbf{1990,} \textsl{17,} 189.

\bibitem{Thompson:91}
Thompson,~M.~A.;\ \ Zerner,~M.~C.;\ \ Fajer,~J. \textit{J. Phys. Chem.}
  \textbf{1991,} \textsl{95,} 5693.

\bibitem{Zhou:97}
Zhou,~H.;\ \ Boxer,~S.~G. \textit{J. Phys. Chem. B} \textbf{1997,}
  \textsl{101,} 5759.

\bibitem{Chang:01}
Chang,~C.~H.;\ \ Hayashi,~M.;\ \ Liang,~K.~K.;\ \ Chang,~R.;\ \ Lin,~S.~H.
  \textit{J. Phys. Chem. B} \textbf{2001,} \textsl{105,} 1216.

\bibitem{Renger:04}
Renger,~T. \textit{Phys. Rev. Lett.} \textbf{2004,} \textsl{93,} 188101.

\bibitem{DMjpca:01}
Matyushov,~D.~V.;\ \ Newton,~M.~D. \textit{J. Phys. Chem. A} \textbf{2001,}
  \textsl{105,} 8516.

\bibitem{DMjcp:00}
Matyushov,~D.~V.;\ \ Voth,~G.~A. \textit{J. Chem. Phys.} \textbf{2000,}
  \textsl{113,} 5413.

\bibitem{Warshel:89}
Warshel,~A.;\ \ Chu,~Z.~T.;\ \ Parson,~W.~W. \textit{Science} \textbf{1989,}
  \textsl{246,} 112.

\bibitem{Treutlein:92}
Treutlein,~H.;\ \ Schulten,~K.;\ \ Br{\"u}nger,~A.~T.;\ \ Karplus,~M.;\ \
  Deisenhofer,~J.;\ \ Michel,~H. \textit{Proc. Natl. Acad. Sci.} \textbf{1992,}
  \textsl{89,} 75.

\bibitem{Gehlen:94}
Gehlen,~J.~N.;\ \ Marchi,~M.;\ \ Chandler,~D. \textit{Science} \textbf{1994,}
  \textsl{263,} 499.

\bibitem{Sterpone:03}
Sterpone,~F.;\ \ Ceccarelli,~M.;\ \ Marchi,~M. \textit{J. Phys. Chem. B}
  \textbf{2003,} \textsl{107,} 11208.

\bibitem{vanDuijnen:98}
van Duijnen,~P.;\ \ Swart,~M. \textit{J. Phys. Chem. A} \textbf{1998,}
  \textsl{102,} 2399.

\bibitem{MarchiParms02}
Ceccarelli,~M.;\ \ Procacci,~P.;\ \ Marchi,~M. \textit{J.\ Comp.\ Chem.}
  \textbf{2003,} \textsl{24,} 129.

\bibitem{Aqvist:93}
{\AA}qvist,~J.;\ \ Warshel,~A. \textit{Chem. Rev.} \textbf{1993,} \textsl{93,}
  2523.

\bibitem{Schmitt:99}
Schmitt,~U.~W.;\ \ Voth,~G.~A. \textit{J.\ Chem.\ Phys.} \textbf{1999,}
  \textsl{111,} 9361.

\bibitem{protRefInd}
Bialek-Bylka,~G.~E.;\ \ Jazurek,~B.;\ \ Dedic,~R.;\ \ Hala,~J.;\ \
  Skrzypczak,~A. \textit{Cell. Mol. Biol. Lett.} \textbf{2003,} \textsl{8,}
  689.

\bibitem{Sasisanker:04}
Sasisanker,~P.;\ \ Oleinikova,~A.;\ \ Weingartner,~H.;\ \ Ravindra,~R.;\ \
  Winter,~R. \textit{Phys. Chem. Chem. Phys.} \textbf{2004,} \textsl{6,}
  1899.

\bibitem{Haffa:02}
Haffa,~A. L.~M.;\ \ Lin,~S.;\ \ Katilius,~E.;\ \ Williams,~J.~C.;\ \
  Taguchi,~A. K.~W.;\ \ Allen,~J.~P.;\ \ Woodbury,~N.~W. \textit{J. Phys. Chem.
  B} \textbf{2002,} \textsl{106,} 7376.

\bibitem{Sulpizi:07}
Sulpizi,~M.;\ \ Raugei,~S.;\ \ VandeVondele,~J.;\ \ Carloni,~P.;\ \ Sprik,~M.
  \textit{J.\ Phys.\ Chem.\ B} \textbf{2007,} \textsl{111,} 3969.

\bibitem{Newton:99}
Newton,~M.~D. \textit{Adv. Chem. Phys.} \textbf{1999,} \textsl{106,} 303.

\bibitem{DMjpca1:06}
Ghorai,~P.~K.;\ \ Matyushov,~D.~V. \textit{J. Phys. Chem. A} \textbf{2006,}
  \textsl{110,} 8857.

\bibitem{DMjpcb1:06}
Ghorai,~P.~K.;\ \ Matyushov,~D.~V. \textit{J. Phys. Chem. B} \textbf{2006,}
  \textsl{110,} 1866.

\bibitem{Ferrand:93}
Ferrand,~M.;\ \ Dianoux,~A.~J.;\ \ Petry,~W.;\ \ Zaccai,~G. \textit{Proc. Natl.
  Acad. Sci.} \textbf{1993,} \textsl{90,} 9668.

\bibitem{Bizzarri:04}
Bizzarri,~A.~R. \textit{J. Phys.: Condens. Matter} \textbf{2004,} \textsl{16,}
  R83.

\bibitem{Kirmaier:88}
Kirmaier,~C.;\ \ Holten,~D.  .   In  \textit{Photosynthetic Bacterial Reaction
  Center: Structure and Dynamics}, Vol.~149; Breton,~J.;\ \ Verm{\
  {\'e}}gio,~A.,\ \ Eds.;  Plenum: New York, 1988.

\bibitem{Huber:98}
Huber,~H.;\ \ Meyer,~M.;\ \ Scheer,~H.;\ \ Zinth,~W.;\ \ Wachtveil,~J.
  \textit{Photosynth. Res.} \textbf{1998,} \textsl{55,} 153.

\bibitem{Renger:07}
Renger,~T.;\ \ Trostmann,~I.;\ \ Theiss,~C.;\ \ Madjet,~M.;\ \ Richter,~M.;\ \
  Paulsen,~H.;\ \ Eichler,~H.;\ \ Knorr,~A.;\ \ Renger,~G. \textit{J. Phys.
  Chem. B} \textbf{2007,} \textsl{111,} 10487.

\bibitem{Vos:99}
Vos,~M.~H.;\ \ Martin,~J.-L. \textit{Biochim. Biophys. Acta} \textbf{1999,}
  \textsl{1411,} 1.

\bibitem{Trissl:01}
Trissl,~H.-W.;\ \ Bernhardt,~K.;\ \ Lapin,~M. \textit{Biochemistry}
  \textbf{2001,} \textsl{40,} 5290.

\bibitem{DMjpcb1:08}
LeBard,~D.~N.;\ \ Matyushov,~D.~V. \textit{J. Phys. Chem. B} \textbf{2008,}  in
  press, arXiv:0709.4282.

\bibitem{DMjpcb3:08}
LeBard,~D.~N.;\ \ Matyushov,~D.~V., unpublished.

\bibitem{DMjpcb3:06}
Kapko,~V.;\ \ Matyushov,~D.~V. \textit{J. Phys. Chem. B} \textbf{2006,}
  \textsl{110,} 13184.

\bibitem{Vos:91}
Vos,~M.~H.;\ \ Lambry,~J.;\ \ Robles,~S.~J.;\ \ Youvan,~D.~C.;\ \ Breton,~J.;\
  \ Martin,~J. \textit{Proc. Natl. Acad. Sci.} \textbf{1991,} \textsl{88,}
  8885.

\bibitem{Du:92}
Du,~M.;\ \ Rosenthal,~S.~J.;\ \ Xie,~X.;\ \ DiMagno,~T.~J.;\ \ Schmidt,~M.;\ \
  Hanson,~D.~K.;\ \ Schiffer,~M.;\ \ Norris,~J.~R.;\ \ Fleming,~G.~R.
  \textit{Proc. Natl. Acad. Sci.} \textbf{1992,} \textsl{89,} 8517.

\bibitem{Chaudhury:07}
Chaudhury,~S.;\ \ Cherayil,~B.~J. \textit{J. Chem. Phys.} \textbf{2007,}
  \textsl{127,} 145103.

\bibitem{Agmon:83}
Agmon,~N.;\ \ Hopfield,~J.~J. \textit{J. Chem. Phys.} \textbf{1983,}
  \textsl{78,} 6947.

\bibitem{Gayathri:96}
Gayathri,~N.;\ \ Bagchi,~B. \textit{J. Phys. Chem.} \textbf{1996,}
  \textsl{100,} 3056.

\bibitem{Okuyama:86}
Okuyama,~S.;\ \ Oxtoby,~D.~W. \textit{J. Chem. Phys.} \textbf{1986,}
  \textsl{84,} 5830.

\bibitem{Hynes:86}
Hynes,~J.~T. \textit{J.\ Phys.\ Chem.} \textbf{1986,} \textsl{90,} 3701.

\bibitem{Walker:92}
Walker,~G.~C.;\ \ {\AA}kesson,~E.;\ \ Johnson,~A.~E.;\ \ Levinger,~N.~E.;\ \
  Barbara,~P.~F. \textit{J. Phys. Chem.} \textbf{1992,} \textsl{96,} 3728.

\bibitem{Abbyad:07}
Abbyad,~P.;\ \ Shi,~X.;\ \ Childs,~W.;\ \ McAnaney,~T.;\ \ Cohen,~B.;\ \
  Boxer,~S. \textit{J. Phys. Chem. B} \textbf{2007,} \textsl{111,} 8269.

\bibitem{Sahu:06}
Sahu,~K.;\ \ Mondal,~S.~K.;\ \ Ghosh,~S.;\ \ Roy,~D.;\ \ Bhattacharyya,~K.
  \textit{J. Chem. Phys.} \textbf{2006,} \textsl{124,} 124909.

\bibitem{LiT:07}
Li,~T.;\ \ Hassanali,~A.;\ \ Kao,~Y.-T.;\ \ Zhong,~D.;\ \ Singer,~S. \textit{J.
  Am. Chem. Soc.} \textbf{2007,} \textsl{129,} 3376.

\bibitem{Markelz:07}
Markelz,~A.~G.;\ \ Knab,~J.~R.;\ \ Chen,~J.~Y.;\ \ He,~Y. \textit{Chem. Phys.
  Lett.} \textbf{2007,} \textsl{442,} 413.

\bibitem{DMjcp2:08}
LeBard,~D.~N.;\ \ Matyushov,~D.~V. \textit{J. Chem. Phys.} \textbf{2008,}  in
  press.

\bibitem{Edens:00}
Edens,~G.~J.;\ \ Gunner,~M.~R.;\ \ Xu,~Q.;\ \ Mauzerall,~D. \textit{J. Am.
  Chem. Soc.} \textbf{2000,} \textsl{122,} 1479.

\bibitem{GunnerDutton:89}
Gunner,~M.~R.;\ \ Dutton,~P.~L. \textit{J. Am. Chem. Soc.} \textbf{1989,}
  \textsl{111,} 3400.

\bibitem{Sumi:01}
Sumi,~H.;\ \ Kakitani,~T. \textit{J. Phys. Chem. B} \textbf{2001,}
  \textsl{105,} 9603.

\bibitem{com:TRP}
Both PHE and TYR also absorb at 280 nm and, in addition, exposure to water
  makes fluorescence decay faster (see, J. A. McCammon, P. G. Wolynes, and M.
  Karplus, \textit{Biochemistry}, \textbf{1979}, \textit{18}, 927). Although
  there are 39 TRP residues used to calculate the TRP Stokes shift dynamics, an
  additional 27 TYR and 58 PHE residues also exist in the wild type, which
  might have contributed to a faster decay of the correlation function compared
  to TRP-only function calculated from MD simulations.

\bibitem{Winsdor:89}
Windsor,~M.~W.;\ \ Menzel,~R. \textit{Chem. Phys. Lett.} \textbf{1989,}
  \textsl{164,} 143.

\bibitem{Cherepy:97}
Cherepy,~N.;\ \ Shreve,~A.;\ \ Moore,~L.;\ \ Boxer,~S.;\ \ Mathies,~R.
  \textit{J. Phys. Chem. B} \textbf{1997,} \textsl{101,} 3250.

\bibitem{parsonWarshel:87}
Warshel,~A.;\ \ Parson,~W.~W. \textit{J. Am. Chem. Soc.} \textbf{1987,}
  \textsl{109,} 6143.

\bibitem{DMjpca:04}
Gupta,~S.;\ \ Matyushov,~D.~V. \textit{J. Phys. Chem. A} \textbf{2004,}
  \textsl{108,} 2087.

\bibitem{ParsonW:04}
Parson,~W.~W.;\ \ Warshel,~A. \textit{Chem. Phys.} \textbf{2004,} \textsl{296,}
  201.

\bibitem{Parson:04}
Parson,~W.~W.;\ \ Warshel,~A. \textit{J. Phys. Chem. B} \textbf{2004,}
  \textsl{108,} 10474.

\bibitem{Beratan:92}
Beratan,~D.~N.;\ \ Betts,~J.~N.;\ \ Onuchic,~J.~N. \textit{J. Phys. Chem.}
  \textbf{1992,} \textsl{96,} 2852.

\bibitem{Skourtis:05}
Skourtis,~S.~S.;\ \ Balabin,~I.~A.;\ \ Kawatsu,~T.;\ \ Beratan,~D.~N.
  \textit{Proc. Natl. Acad. Sci.} \textbf{2005,} \textsl{102,} 3552.

\bibitem{Gray:05}
Gray,~H.~B.;\ \ Winkler,~J.~R. \textit{Proc. Natl. Acad. Sci.} \textbf{2005,}
  \textsl{102,} 3534.

\bibitem{Moser:06}
Moser,~C.~C.;\ \ Page,~C.~C.;\ \ Dutton,~P.~L. \textit{Phil. Trans. R. Soc.
  London B} \textbf{2006,} \textsl{361,} 1295.

\bibitem{Jasaitis:07}
Jasaitis,~A.;\ \ Johansson,~M.~P.;\ \ Wilkstr{\"o}m,~M.;\ \ Vos,~M.~H.;\ \
  Verhovsky,~M.~I. \textit{Proc. Natl. Acad. Sci.} \textbf{2007,} \textsl{104,}
  20811.

\bibitem{Beratan:08}
Beratan,~D.~N.;\ \ Balabin,~I.~A. \textit{Proc. Natl. Acad. Sci.}
  \textbf{2008,} \textsl{105,} 403.

\bibitem{Goldsmith:05}
Goldsmith,~R.~H.;\ \ Sinks,~L.~E.;\ \ Kelley,~R.~F.;\ \ Betzen,~L.~J.;\ \
  Liu,~W.;\ \ Weiss,~E.~A.;\ \ Ratner,~M.~A.;\ \ Wasielewski,~M.~R.
  \textit{Proc. Natl. Acad. Sci.} \textbf{2005,} \textsl{102,} 3540.

\bibitem{Lewis:001}
Lewis,~F.~D.;\ \ Letsinger,~R.~L.;\ \ Wasielewski,~M.~R. \textit{Acc. Chem.
  Res.} \textbf{2001,} \textsl{34,} 159.

\bibitem{Skov:98}
Skov,~L.;\ \ Pascher,~T.;\ \ Winkler,~J.;\ \ Gray,~H. \textit{J. Am. Chem.
  Soc.} \textbf{1998,} \textsl{120,} 1102.

\bibitem{Cascella:06}
Cascella,~M.;\ \ Magistrato,~A.;\ \ Tavernelli,~I.;\ \ Carloni,~P.;\ \
  Rothlisberger,~U. \textit{Proc. Natl. Acad. Sci.} \textbf{2006,}
  \textsl{103,} 19641.

\bibitem{Blumberger:06}
Blumberger,~J.;\ \ Klein,~M.~L. \textit{J. Am. Chem. Soc.} \textbf{2006,}
  \textsl{128,} 13854.

\bibitem{shake77}
Ryckaert,~J.-P.;\ \ Ciccotti,~G.;\ \ Berendsen,~H. J.~C. \textit{J.\ Comp.\
  Phys.} \textbf{1977,} \textsl{23,} 327.

\bibitem{pme95}
Essmann,~U.;\ \ Perera,~L.;\ \ Berkowitz,~M.~L.;\ \ Darden,~T.;\ \ Lee,~H.;\ \
  Pedersen,~L.~G. \textit{J. Chem. Phys.} \textbf{1995,} \textsl{103,}
  8577.

\bibitem{gamess}
Schmidt,~M.~W.;\ \ Baldridge,~K.;\ \ Boatz,~J.~A.;\ \ Elbert,~S.~T.;\ \
  Gordon,~M.~S.;\ \ Jensen,~J.~H.;\ \ Koseki,~S.;\ \ Matsunaga,~N.;\ \
  Nguyen,~K.~A.;\ \ Su,~S.~J.;\ \ Windus,~T.~L.;\ \ Dupuis,~M.;\ \
  Montgomery,~J.~A. \textit{J. Comput. Chem.} \textbf{1993,} \textsl{14,} 1347.

\bibitem{Johnson:02}
Johnson,~E.~T.;\ \ M{\"u}h,~F.;\ \ Nabedryk,~E.;\ \ Williams,~J.~C.;\ \
  Allen,~J.~P.;\ \ Lubitz,~W.;\ \ Breton,~J.;\ \ Parson,~W.~W. \textit{J. Phys.
  Chem. B} \textbf{2002,} \textsl{106,} 11859.

\bibitem{amberFF03}
Duan,~Y.;\ \ Wu,~C.;\ \ Chowdhury,~S.;\ \ Lee,~M.~C.;\ \ Xiong,~G.;\ \
  Zhang,~W.;\ \ Yang,~R.;\ \ Cieplak,~P.;\ \ Luo,~R.;\ \ Lee,~T.;\ \
  Caldwell,~J.;\ \ Wang,~J.;\ \ Kollman,~P. \textit{J.\ Comp.\ Chem.}
  \textbf{2003,} \textsl{24,} 1999.

\bibitem{tip3p:83}
Jorgensen,~W.~L.;\ \ Chandrasekhar,~J.;\ \ Madura,~J.~D.;\ \ Impey,~R.~W.;\ \
  Klein,~M.~L. \textit{J. Chem. Phys.} \textbf{1983,} \textsl{79,} 926.

\bibitem{BixonJortner:89}
Bixon,~M.;\ \ Jortner,~J.;\ \ Michel-Beyerle,~M.~E.;\ \ Ogrodnik,~A.
  \textit{Biochim. Biophys. Acta} \textbf{1989,} \textsl{977,} 273.

\end{thebibliography}

\end{document}